\pdfoutput = 1
\documentclass[aps,prl,showpacs,twocolumn,longbibliography]{revtex4-1}

\usepackage[T1]{fontenc}
\usepackage[latin9]{inputenc}
\usepackage{float}

\usepackage{amsmath,amssymb,amsfonts}

\usepackage{graphicx}

\usepackage{esint}
\usepackage[normalem]{ulem}

\usepackage[colorlinks,bookmarks=false,citecolor=blue,linkcolor=blue,urlcolor=blue,hyperfootnotes=true]{hyperref}

\newcommand{\pf}{2{\bf {p}}_{\text{F}}}

\makeatother

\begin{document}

\title{Charge orders, magnetism and pairings in the cuprate superconductors}

\date{December 11, 2015}

\author{T.\ Kloss$^{1}$, X.\ Montiel$^{1}$, V. S. de Carvalho$^{2}$,
H. Freire$^{2}$, C.\ P\'epin$^{1}$}

\affiliation{1. IPhT, L'Orme des Merisiers, CEA-Saclay, 91191 Gif-sur-Yvette,
France }

\affiliation{2. Instituto de F\'isica, Universidade Federal de Goi\'as, 74.001-970,
Goi\^{a}nia-GO, Brazil}
\begin{abstract}
We review the recent developments in the field of cuprate superconductors
with the special focus on the recently observed charge order in the
underdoped compounds. We introduce new theoretical developments following
the study of the antiferromagnetic (AF) quantum critical point (QCP)
in two dimensions, in which preemptive orders in the charge and superconducting
(SC) sectors emerged, that are in turn related by an SU(2) symmetry. We consider
the implications of this proliferation of orders in the underdoped
region, and provide a study of the type of fluctuations which characterize
the SU(2) symmetry. We identify an intermediate energy scale where
the SU(2) pairing fluctuations are dominant and argue that they are unstable
towards the formation of a Resonant Peierls Excitonic (RPE) state
at the pseudogap (PG) temperature $T^{*}$. We discuss the implications
of this scenario for a few key experiments. 
\end{abstract}
\maketitle

\section{Introduction}

The last decade has seen a strong revival of interest in cuprate superconductors,
with the observation of charge orders in the underdoped regime of
these materials. Maybe the starting point of this intense
period of investigation was the observation by STM of checkerboard-type
patterns inside the vortices in Bi-2212 \cite{Hoffman02,Fujita14}.
Subsequent studies with Fermi surface reconstruction showed that this
feature was generic \cite{Vershinin:2004gk,daSilvaNeto:2014vy} (also
verified in Bi-2201 \cite{He14,Wise08}) and that the charge patterns
corresponded to two axial wave vectors $\left(0,Q_{y}\right)$ and
$\left(Q_{x},0\right)$, incommensurate with the lattice periodicity,
and the magnitude of the wave vectors decreases with oxygen-doping.
The charge excitation was also found to be non-dispersive in temperature,
and correlated with the ``hot-spots'' - the points of the Fermi
surface where the AF zone boundary is intersected. The picture refined itself
a bit later, and we now believe the charge order emerges at the tip
of the Fermi arcs \cite{Hamidian15,Hamidian15a}. The study of charge
order in underdoped cuprates stayed in a \emph{status quo} until the observation
of quantum oscillations (QO) under a strong magnetic field in YBCO
\cite{Doiron-Leyraud07,LeBoeuf07}. This result pointed directly to
a reconstruction of the Fermi surface induced by magnetic field and
received several explanations in terms of stripe and charge patterns
until the link was made with the bi-axial charge order observed by
STM \cite{Doiron-Leyraud13,Sebastian10,Sebastian12,Cyr15a}.
In particular, models for the reconstruction of the Fermi surface
involved charge ordering with bi-axial wave vectors similar to those
unveiled by STM. A subsequent Nuclear Magnetic Resonance (NMR) study
finally found some charge splitting under a magnetic field $B\geq17$T,
which brought the final confirmation that charge order under a finite
magnetic field is coherent, static and long-ranged \cite{Wu11,Wu13a,Wu:2015bt}.
The field versus temperature phase diagram was later completed by ultrasound experiments, which showed evidence for a flat transition line
at $B_{c}=17$T \cite{LeBoeuf13}. For $B\leq B_{c}$, YBCO is a $d$-wave
superconductor. The increase of the magnetic field then creates vortices
whose cores show the typical charge ordering \cite{Wu13a}. For $B\geq B_{c}$
YBCO shows long range charge order with a typical ordering temperature
remarkably similar in magnitude with the SC $T_{c}$. In the PG regime
at $B=0,$ both hard x-ray \cite{Chang12,Blackburn13a} and soft
x-rays \cite{Ghiringhelli12,LeTacon11,Blanco-Canosa13,Blanco-Canosa14}
study showed the presence of a sizable short range excitation at the
bi-axial wave vectors. A softening of the phonon spectrum has been
observed in the PG phase, while a softening at the charge order wave
vectors occurs below $T_{c}$ \cite{Blackburn13b,LeTacon14}. Note
that a recent state-of-the-art x-ray experiment at $B=17$T
showed that the charge order becomes uni-axial and tri-dimensional
at high field \cite{Chang16}. One preliminary conclusion that one can
infer from these experiments is that the charge and superconducting
sectors are of the same order of magnitude in cuprate superconductors.
We will use this observation later when we introduces the emergent SU(2) rotations that appear
between these two sectors.

We turn now to one of the most enduring mysteries of cuprate superconductors,
the pseudogap (PG) regime. The PG phase was observed in 1989 by NMR experiments \cite{Alloul89,Warren89}, where a gradual
drop in the Knight-shift was observed at a crossover temperature
$T^{*}$. This gap was attributed to a loss of density in the electronic
carriers, and it was shown to decrease when the oxygen-doping increases
but no obvious symmetry breaking was associated with this phase transition.
We focus here on a few properties of the PG phase which we will use
later in the SU(2)-interpretation of the experiments. The first remark
that one can make is that the PG is an extremely robust feature of
the phase diagram. It seems insensitive to disorder \cite{Alloul:2010ko,RullierAlbenque:2000fl}
and magnetic field \cite{RullierAlbenque:2007bm} and is closely associated
to a regime of linear-in-$T$ resistivity on its right hand side \cite{Hussey:2008tw,Hussey:2011kp,husseycupratescriticality}.
One other very striking observation in the PG regime detected by
angle-resolved photoemission (ARPES), is the formation of Fermi arcs,
instead of a closed-contour Fermi surface \cite{Campuzano98,Campuzano99,Chatterjee06,Kanigel07,Shen:2004ek,Shen:2005ir}.
Recently, a momentum scale of similar magnitude as the one observed
in the CDW was associated to the opening of the PG in the
anti-nodal region of the Brillouin zone (BZ) \cite{Shen05,He11,Yoshida:2012kh}, and
led to an interpretation in terms of a pair-density-wave (PDW)
\cite{Lee14,Wang15b}- or a finite momentum superconducting state Fulde-Ferrell-Larkin-Ovchinnikov
(FFLO) \cite{Fulde:1964dq,Larkin65}. Coherent neutron scattering showed a
$\mathbf{Q}=0$ signal \cite{Fauque06,Li08,Bourges11,Sidis13,Mangin_Thro14,ManginThro:2015fg}, which was interpreted
in terms of intra-unit-cell loop currents \cite{Varma06,AjiVarma07,AjiVarma07a}.
Although a $\mathbf{Q}=0$ phase is unable to open a gap in the electronic
density of states, the loop-current line surprisingly follows the
$T^{*}$-line. Note that NMR \cite{Roos2011,Mounce2013} and $\mu$SR \cite{MacDougall08,Sonier2009} techniques were not able to detect such loop current. An explanation could be the longer time scale of local probes  ($\approx 10^{-8}-10^{-6}s$) compared with the INS time scale ($\approx 10^{-11}s$). 
At lower temperature, a Kerr effect signal has been
reported, hinting at a breaking of time-reversal (TR) symmetry inside
the PG \cite{Xia08}. This last observation is widely discussed by
the community, but it is necessarily related to the $\mathbf{Q}=0$ loop currents
\cite{Aji:2013eo,Yakovenko:2015fd}. The inelastic neutron scattering
(INS) is also interesting for revealing collective modes of the system.
A resonance at $41$meV was found in YBCO the early days of cuprate
superconductivity \cite{Rossat} and at similar energies in other compounds \cite{Bourges2005,Sidis2001,Fong99,He02}. It was first believed
that this collective excitation existed only in the SC phase, where
it has a typical ``hour-glass'' shape centered around $41$ meV
at $\mathbf{Q}=\left(\pi,\pi\right)$, as a function of energy and wave-vector.
It was later shown that the resonance exists as well in the PG phase
above $T_{c}$, where it is still centered around $41$\,meV, but
shows a typical ``Y''-shape with a long energy-extension at $\mathbf{Q}=\left(\pi,\pi\right)$\cite{Bourges19052000,Hinkov04,Hinkov07,Hayden04}. Many theoretical approaches have been
invoked to describe the resonance below the SC transition \cite{Norman07,Eschrig:2000bf,Eremin:2005ba,Demler95}.
This observation of the resonance around similar typical energies
in the SC and PG phases, however, has never received a theoretical
description, and constrains theories of the PG to keep some reminiscence
of the SC phase. The neutron resonance was also observed in mono-layer
tetragonal compounds (Hg-1201), where the long energy extension at
$\mathbf{Q}=\left(\pi,\pi\right)$ persists below $T_{c}$ \cite{Tabis14}.

Collective modes of a material
give useful insights to probe symmetries of an effective model. 
One example is a resonance observed in the Raman $A_{1g}$ channel,  that appears at energies very similar to the ones where
a collective mode was observed by INS \cite{Gallais04,Blanc:2009vo,Sacuto2015TS}.
Raman scattering typically probes the symmetries of the Fermi surface
and the presence of ``two gaps'' in the underdoped regime of the
cuprates was observed below $T_{c}$ \cite{LeTacon06,Civelli:2008vr,Blanc:2010tm}.
This fact was corroborated in a series of ARPES experiments on
BSCO from which the gap velocity $v_{\Delta}$ at the nodes was extracted
and shown to differ from the Fermi velocity. Three regions in the
phase diagram were identified \cite{Vishik12}. Starting from the
over-doped region and decreasing the doping, $v_{\Delta}$ is shown
to first increase then to reach a plateau in the underdoped region -down
to dopings of the order of 5 \%, and after that it drops at lower dopings when the
system gets close to the insulating Mott-transition. The key question
associated with the PG phase is whether it is a ``strong-coupling''
phenomenon, emerging as a direct consequence of the Mott transition
\cite{Anderson87,Georges96,Katanin04,Sordi:2012jc,Gull:2013hh}, or whether
it is a a very unusual collective phenomenon which is sensitive to
other peculiarities of the physics of the cuprates, like its low dimensionality,
the antiferromagnetic fluctuations or its fermiology \cite{LeHur:2009iw,Rice12,Chowdhury:2015gx,Kloss15a}.
In this work, we argue that the key to explain the mystery of the
PG phase resides in an underlying emergent SU(2) symmetry, which produces SU(2)
pairing fluctuations at intermediate energy scales. These fluctuations
are in turn unstable toward the formation of a new kind of excitonic
state, the (RPE) state, which is responsible for gapping out the Fermi
surface in the anti-nodal region of the BZ \cite{Kloss15a}.

The paper is organized as follows: In section \ref{sectionSU2}, 
we present the basics of the emergent symmetry model with SU(2) symmetry.  
Section \ref{sectionfluc} discusses the competition between the U(1) and SU(2) paring fluctuations in the framework of the non linear $\sigma$ model. 
In particular, we propose to explain the PG state as a new type of charge order: the Resonant Peierls Excitonic (RPE) state coming from the SU(2) fluctuations. 
We also demonstrate that the CDW state is a secondary instability produced by U(1) fluctuations mediated by a Leggett mode. 
In section \ref{sectionDiscussion}, we discuss the possible experimental evidence of this phase before to conclude in section \ref{conclusion}.

\section{The emergent SU(2) symmetry}
\label{sectionSU2}
The concept of emergent symmetry in the context of the cuprate superconductors
can be traced back to the work of Yang and later Zhang \cite{Yang89,Yang:1990cf}
where a representation with pseudo-spin operators was introduced which
rotated the $d$-wave SC state onto a $d$-wave bi-partite charge
order. The lowering and raising pseudo-spin operators $\eta^{+},$
$\eta^{-}=\left(\eta^{+}\right)^{\dagger}$, and $\eta_{z}$, which
follow from the definition 
\begin{subequations} \label{eq:1} 
\begin{align}
\eta^{+} & =\sum_{\mathbf{k}}c_{\mathbf{k}\uparrow}^{\dagger}c_{-\mathbf{\mathbf{k}+Q}\downarrow}^{\dagger}\\
\eta_{z} & =\sum_{\mathbf{k}}\left(c_{\mathbf{k}\uparrow}^{\dagger}c_{\mathbf{k}\uparrow}+c_{\mathbf{k+Q}\downarrow}^{\dagger}c_{\mathbf{k+Q}\downarrow}-1\right).
\end{align}
\end{subequations} The operators (\ref{eq:1}) form an SU(2) Lie
algebra. Noticeably, the $\eta$-pairing stat -- which is equivalent
to finite center of mass pairing of vector $\mathbf{Q}=\left(\pi,\pi\right)$,
or FFLO state -- is an eigenstate of the Hubbard Hamiltonian, both for
positive and negative U. The simplest irreducible representation for
the pseudo-spin is the triplet vector $\Delta_{m},$ with $m=\left\{ -1,0,1\right\} $
defined as \begin{subequations} \label{eq:1-1} 
\begin{align}
\Delta_{1} & =-\frac{1}{\sqrt{2}}\sum_{\mathbf{k}}c_{\mathbf{k}\uparrow}^{\dagger}c_{-\mathbf{k}\downarrow}^{\dagger},\\
\Delta_{0} & =\frac{1}{2}\sum_{\mathbf{k},\sigma}c_{\mathbf{k}\sigma}^{\dagger}c_{\mathbf{k+Q}\sigma},\\
\Delta_{-1} & =-\Delta_{1}^{\dagger},
\end{align}
\end{subequations} which represents the two conjugated s-wave SC
states $\left( \Delta_{-1} , \Delta_{1} \right)$
and the charge ordering state $ \Delta_{0}$. The SU(2)
pseudo-spin operators (\ref{eq:1}) rotate each component of the multiplet
(\ref{eq:1-1}) into one another in the standard way ($l$ is the
rank of the irreducible representation Eq.\ (\ref{eq:1-1}), here
$l=1$) \begin{subequations} \label{eq:3} 
\begin{align}
\left[\eta^{\pm},\Delta_{m}\right] & =\sqrt{l\left(l+1\right)-m\left(m\pm1\right)}\Delta_{m\pm1},\\
\left[\eta_{z},\Delta_{m}\right] & =m\Delta_{m}.
\end{align}
\end{subequations} The effective theory describing the pseudo-spin
symmetry is the $SO\left(4\right)$ {[}$SO\left(4\right)=(SU\left(2\right)\times SU\left(2\right))/Z_{2}${]}
non-linear $\sigma$-model which excites thermally from the SC state
to the ordered state. This model describes transitions from one
state to the other within the generic framework of ``spin-flop''
transitions. In the case above one has a pseudo spin-flop from the
s-wave SC to the CDW states, whereas the standard spin-flop transition from easy axis to easy plane belongs to the SO(3) group \cite{Zhang:1997ew}.
The concept of SU(2)-symmetry was used later on in an effective theory
of the PG leading to a rotation from the $d$-wave superconductor to the d-density
wave state \cite{Nayak00}. Here the generators of the symmetry are
simply $i\eta^{+}$, $i\eta^{-}$ and $\eta^{z}$ and the effective
theory is the $O\left(4\right)$ non-linear $\sigma$-model. Let us
mention a similar rotation between the nematic $d$-wave bond order $ \Delta_{nem} =\frac{1}{2}$
$\sum_{\mathbf{k}\sigma}d_{\mathbf{k}}  c_{\mathbf{k}\sigma}^{\dagger}c_{\mathbf{k}\sigma}  $
and $d$-wave states $ \Delta_{dsc}^{+} =-\frac{1}{\sqrt{2}}\sum_{\mathbf{k}}d_{\mathbf{k}}  c_{\mathbf{k}\uparrow}^{\dagger}c_{\mathbf{k}\downarrow}^{\dagger}  $,
and $ \Delta_{dsc}^{-} =\frac{1}{\sqrt{2}}\sum_{\mathbf{k}}d_{\mathbf{k}}  c_{\mathbf{k}\downarrow}c_{\mathbf{k}\uparrow}  $
where $d_{\mathbf{k}}=\cos k_{x}-\cos k_{y}$ \cite{Kee:2008gw}.
The pseudo-spin generators in this case take the form $L^{+}$, $L^{-}=\left(L^{+}\right)^{\dagger}$,
$L_{0}$ with 
\begin{equation}
L^{+}=\sum_{\mathbf{k}}c_{\mathbf{k}\uparrow}^{\dagger}c_{-\mathbf{k}\downarrow}^{\dagger},\qquad L_{0}=\sum_{\mathbf{k}\sigma}\left(c_{\mathbf{k}\sigma}^{\dagger}c_{\mathbf{k}\sigma}-1\right).\label{eq:1-2}
\end{equation}
Note that the chemical potential couples to the generator $\eta^{z}$
(or $L_{0}$) and thus a finite chemical potential breaks the SU(2)
symmetry in favor of the SC state.

\begin{figure}[tb]
\includegraphics[width=70mm]{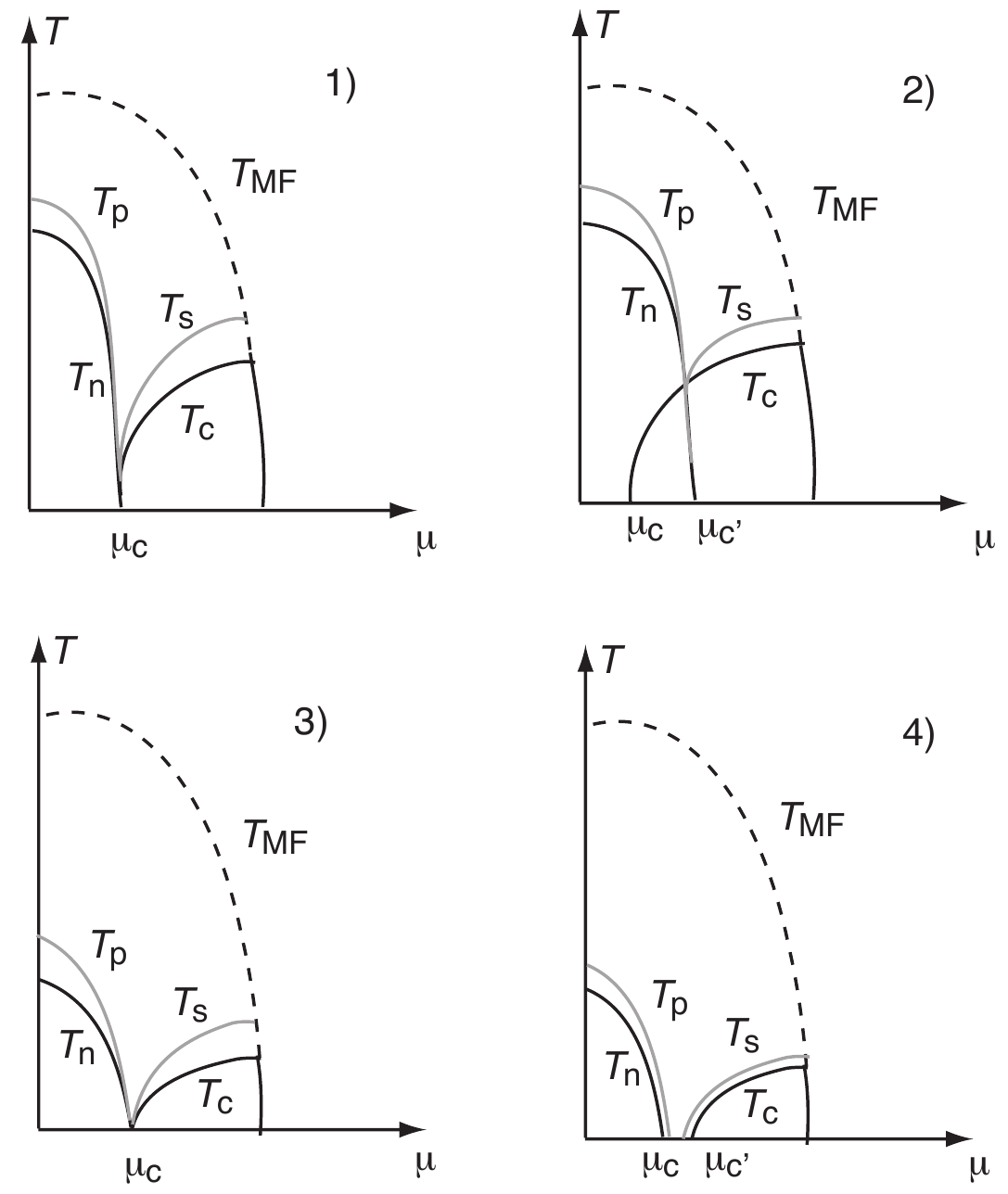} \vspace{-0.5ex}
 \caption{\label{fig:so5} (Color online) Schematic phase diagram of the SO(5)
model \cite{zhang97}. %$T_{MF}$, $T_N$  and $T_c$  are the  electron singlet paring, N\'eel,%  and  the SC critical temperatures. 
Four types of scenarios are discussed in Ref.\ \cite{zhang97}: 1)
a direct first order transition with a bi-critical point, 2) two second
order transitions with an intermediate coexistence regime, 3) one
single second order transition terminating at a QCP at zero temperature
and 4) two second order transitions with a quantum disordered phase.
Although the SO(5) symmetry is broken in scenario 1), 2) and 3) at
zero temperature, thermal fluctuations lead to a restoration below
the mean-field critical temperature $T_{MF}$. Adapted from Ref.\ \cite{zhang97}.}
\end{figure}

Another rotation, this time from the SC state towards the AF state,
was introduced early on and became famous as the SO(5) theory \cite{Zhang:1997ew,Zhang:1999fe,Demler04}
. The SO(5) theory is the one of a non-linear $\sigma$-model which
operates on a five state ``superspin'' $\left(\begin{array}{ccccc}
n_{1}, & n_{2}, & n_{3}, & n_{4}, & n_{5}\end{array}\right)$- two SC states $\left(\begin{array}{cc}
n_{1}=\Delta_{s}, & n_{5}=\Delta_{s}^{\dagger}\end{array}\right)$ and three AF vectors $\left(\begin{array}{ccc}
n_{2}=s^{+}, & n_{3}=s^{-}, & n_{4}=s^{z}\end{array}\right)$ \cite{Zhang:1997ew}. The superspin $n_{a}$ is a vector representation
of the SO(5) algebra. The SO(5) theory was based on the idea that
both the SC and the AF states are key players of the physics of these
compounds and are close enough in energy so that in between their
respective phase transition an SO(5)-symmetric state is found where
SC and AFM are undistinguishable. This phase was naturally associated
with the PG of the cuprates. A typical SO(5) non-linear $\sigma$-model
was introduced to describe the effective physics of the system, and
four typical phase diagrams were derived which are depicted in Fig.
\ref{fig:so5}. The mechanism favoring one of the states in the non-linear $\sigma$-model can be understood as a spin-flop transition-
also called ``super spin flop'' transition for the SO(5) symmetries.
As mentioned above, one gets a very accurate picture by thinking of
the spin-flop transition of the antiferromagnetic state in a uniform
magnetic field $\mathbf{B}$ along the easy$z$-axis \cite{Chakravarty:1988uu,Chakravarty89}.
The magnetic field creates an easy plane $xy$, so that at a critical
value of the field, the N\'eel wave vector changes its orientation abruptly
from the $z$-axis to the $xy$-plane. Hence although in each of the
above cases the symmetries are different, the underlying physics is
as simple as the one on a spin-flop transition. The four typical phase
diagrams show the various phases as a function of temperature and
an external parameter which breaks the symmetry and are depicted in
Fig. \ref{fig:so5}. They correspond to the cases
where: 1) a first order transition between the two states terminates
at a bi-critical point; 2) there is a coexistence phase between
the two orders; 3) the transition between the two orders terminates
at $T=0$ at a quantum critical point (QCP) or 4) the two orders are
disconnected. In cases 1), 2) and 3), although the symmetry is broken
at zero temperature, thermal fluctuations restore the SO(5)- or SU(2)-
symmetry, leading to an invariant phase under the mean field transition
$T_{MF}$. In the case of the SO(5)-symmetry, the typical phase diagram
of the cuprates has the shape depicted in case 4), where the two orders
are disconnected from one another, and this situation leads 
to a splitting of the big group to the $SO(3)\times U(1)$ subgroups
describing fluctuations associated to each order separately. Furthermore,
the SO(5) symmetry led to the prediction of a giant proximity effect
in a SC-AF-SC junction \cite{Demler:1998iw,denHertog:1999dk}. Unfortunately, 
this proximity effect was never verified experimentally \cite{Bozovic:2003hx}.
One reason invoked here was that the SC state is by nature itinerant
while the AF state is an insulator in those compounds. Hence the
typical energy difference between those two states is big, of the
order of the Coulomb U.

In the present work, we revive the concept of emergent symmetry, with an 
the SU(2) symmetry which rotates from a $d$-wave SC state to an incommensurate
$d$-wave charge order. The pseudo-spin generators have the same form
as depicted in Eqn. (\ref{eq:1}), with the exception that the wave
vector $\mathbf{Q=Q_{0}}$ is now the charge order wave vector and
is not necessarily commensurate with the lattice. The $l=1$ irreducible
representation is given by the $d$-wave version of Eqn. (\ref{eq:1-1})
with namely $\Delta_{0}=\chi_{CDW}$, $\Delta_{1}=\Delta_{dsc}^{\dagger}$and
$\Delta_{-1}=-\Delta_{1}^{\dagger},$ namely \begin{subequations}
\label{eq:5} 
\begin{align}
\Delta_{1} & =-\frac{1}{\sqrt{2}}\sum_{\mathbf{k}}d_{\mathbf{k}}c_{\mathbf{k}\uparrow}^{\dagger}c_{-\mathbf{k}\downarrow}^{\dagger},\\
\Delta_{0} & =\frac{1}{2}\sum_{\mathbf{k},\sigma}d_{\mathbf{k}}c_{\mathbf{k}\sigma}^{\dagger}c_{\mathbf{k+Q}\sigma},\\
\Delta_{-1} & =\frac{1}{\sqrt{2}}\sum_{\mathbf{k}}d_{\mathbf{k}}c_{\mathbf{k}\downarrow}c_{-\mathbf{k}\uparrow}.
\end{align}
\end{subequations}

\begin{figure}[tb]
\includegraphics[width=50mm]{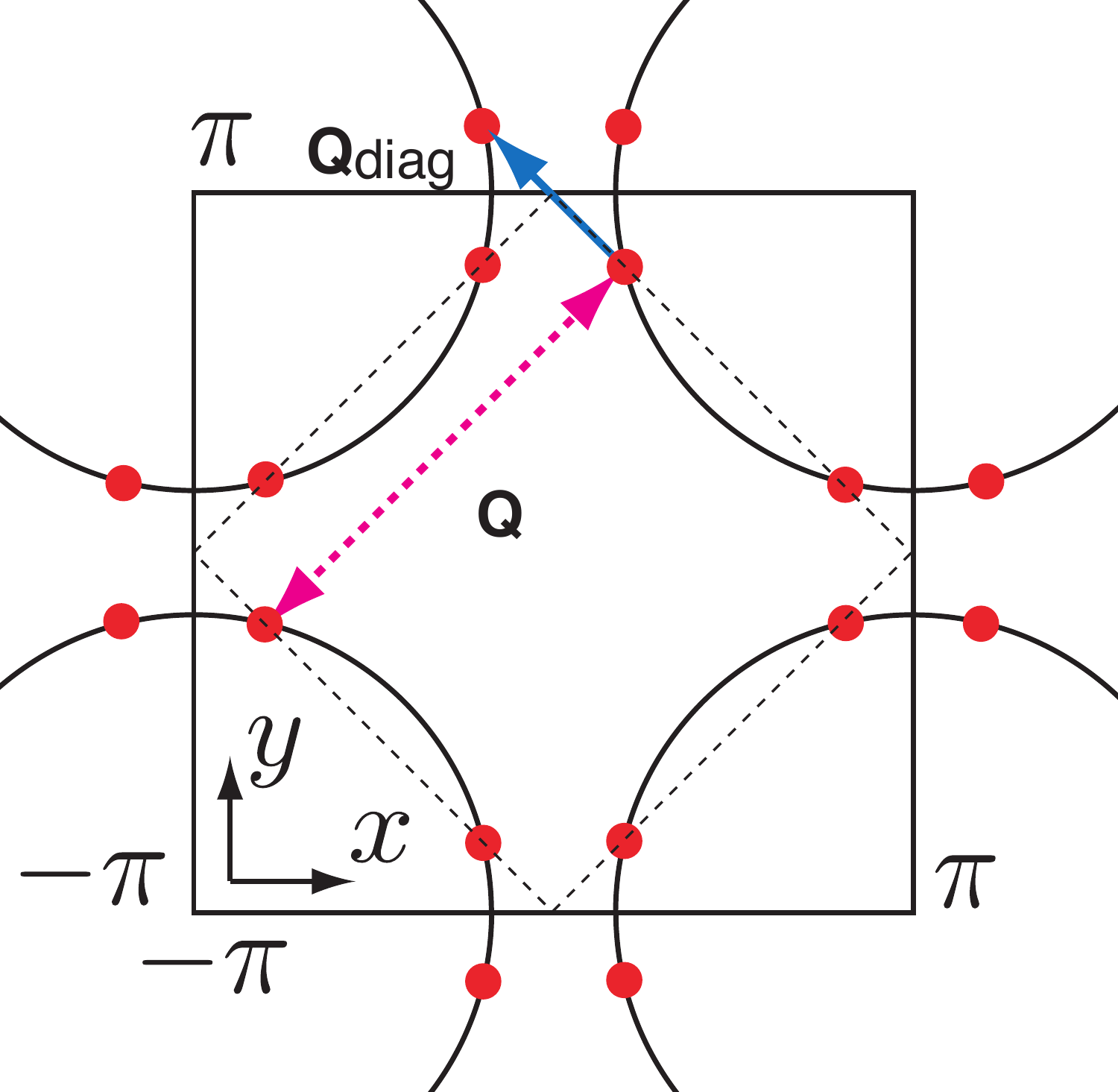} \vspace{-0.5ex}
 \caption{\label{fig:ehsmodel} (Color online) Schematic representation of the
first Brillouin zone with the Fermi surface of cuprate superconductors.
Hotspots are located at the intersection points of the AFM zone boundary
with the Fermi surface. The diagonal coupling vector $\mathbf{Q}_{diag}$ between two hotspots is shown in blue and the AFM ordering vector $\bf{Q}$ in magenta.}
\end{figure}

\begin{figure}[tb]
\includegraphics[width=70mm]{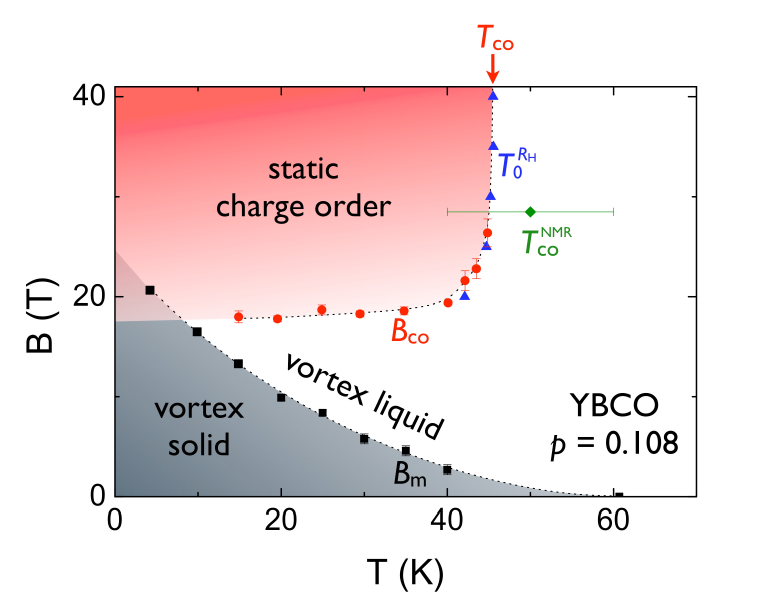} \vspace{-1ex}
 \caption{\label{fig:tb} (Color online) Experimental $B$-$T$ phase
diagram from sound velocity measurements in YBCO from Ref.\ \cite{LeBoeuf13}.
The ``pseudo spin-flop'' is visible from the SC to CDW transition
beyond a critical field $B_{c}\simeq18\,$T.}
\end{figure}

The CDW ordering wave vector could be the axial CDW wave vector observed
through many recent experiments ( STM, Quantum oscillations, X-rays,
ARPES) or it could be another wave vector carefully chosen so that
the SU(2) symmetry is fully respected. As it turns out, the Eight
Hot Spots (EHS) model depicted in Fig.\ \ref{fig:ehsmodel} provides
an exact realization of such a symmetry, as was first mentioned in
Ref.\cite{Metlitski10,Metlitski10a,Efetov13}. 
This model is a simplified
version of the spin-fermion model, which describes the vicinity of
an AF QCP within a metallic substrate \cite{Abanov00}. At this point
it is useful to recall that the spin-fermion model played an important
role at the beginning of the theoretical investigation on the cuprates
\cite{Barzykin:1995fr,Scalapino:2006uw,Monthoux:2007ha}.
For a recent link to strongly correlated systems note also Ref.\ \cite{Ferraz:2015voa}. Two different
views were (and are still) competing for the understanding of the
phase diagram of these compounds. Observing that the SC phase is close
to a Mott insulator, a first group of theoreticians consider that
the system is fundamentally strongly correlated, namely that the Coulomb
energy $U=1$eV is affecting the qualitative behavior down to very
low temperatures \cite{Anderson87,Anderson04,Gull:2013hh}. This viewpoint has been extensively developed around the resonating valence
bond (RVB) suggestion made by Anderson as early as 1987 \cite{Anderson87},
and now explored via extensive numerical calculations which can capture
the strongly interacting behavior \cite{Gull:2013hh,Sordi2012}. Another
part of the physics community defends the viewpoint that the Mott
transition has a strong qualitative influence up to 6-7\% doping,
beyond which the physics of the system is mainly driven by the presence
of AF fluctuations \cite{Pines2002}. Proponents of this viewpoint 
hence start the theoretical investigation with the spin-fermion (SF)
model which couples conduction electrons to AF paramagnon modes $\Phi=\left(\Phi^{x},\Phi^{y},\Phi^{z}\right)$
on the brink of criticality with the propagator 
\begin{align}
\left\langle \Phi_{\omega,\mathbf{q}}^{\alpha}\Phi_{-\omega,-\mathbf{q}}^{\beta}\right\rangle  & =\frac{\delta_{\alpha\beta}}{c^{-2}\omega^{2}+\left(\mathbf{q-Q}\right)^{2}+\xi_{AF}^{-2}},
\end{align}
where \textbf{Q} is the AF wave vector and $\xi_{AF}^{-1}$ is the
effective mass of the paramagnons which defines the distance to the
QCP. Note, however, that in the present theory the SF model can be
considered as an effective theory for which the SU(2) symmetry is
approximately verified -in the case of hot regions \cite{Kloss15},
or exactly verified in the case of the EHS model \cite{Metlitski10,Efetov13}.
Although the proximity to the AF QCP may not be verified in the cuprates,
we believe that the concept of emergent $SU(2)$ symmetry is robust
and will remain when dust settles down. The conduction electrons have
the kinetic energy $H_{K}=\sum_{k,\sigma}c_{k,\sigma}^{\dagger}\epsilon_{k}c_{k,\sigma}$
and interact with the paramagnons through a simple spin-spin interaction
term $H_{int}=J\sum_{i}\mathbf{\Phi}_{i}\cdot c_{i}^{\dagger}\mathbf{\sigma}c_{i}$.
In the EHS model, a further simplification is implemented with the
reduction of the Fermi surface to eight ``hot spots'' which are
the points at $T=0$ where the electrons scatter through the AF $\Phi$-modes.
When the electron dispersion $\epsilon_{k}\simeq v_{hs}k$ is linearized
at the hot spots, the model possesses an exact $SU(2)$ symmetry defined
by the operators of Eqn. (\ref{eq:1}) but with $Q=Q_{diag}$, being
the diagonal wave vector depicted in Fig.\ \ref{fig:ehsmodel}. This
model was further studied in Ref.\cite{Efetov13} and an $SU(2)$
precursor of the AF state was found, where quadrupolar density wave
(QDW) with diagonal wave vector, which is equivalent to a $d$-wave CDW
with diagonal wave vector, is degenerate with the $d$-wave SC state.
This new state can be understood as a kind of non-Abelian superconductor
with order parameter $\hat{b}$ that, instead of having a U(1) phase,
has an $SU(2)$ unitary matrix fluctuating between the charge and
SC sector \cite{Efetov13,Kloss15} 
\begin{align}
\hat{b} & =b\left(\begin{array}{cc}
\Delta_{CDW} & \Delta_{SC}\\
-\Delta_{SC}^{*} & \Delta_{CDW}^{*}
\end{array}\right)_{SU\left(2\right)}
\end{align}
subject to the constraint $\left|\Delta_{CDW}\right|^{2}+\left|\Delta_{SC}\right|^{2}=1$.
Within the framework of the EHS model, and the related $O\left(4\right)$
non linear $\sigma$-model, several experimental findings were successfully
addressed \cite{Meier13,Einenkel14,Hayward14}. The general picture
follows closely the ideas expressed in the $SO(5)$ theory, which
are valid for all theories of emergent symmetries. A small curvature
term in the electron dispersion breaks the symmetry in favor of the
SC state. Hence at $T=0$ the system is a superconductor. Once the
temperature is raised, thermal fluctuations then excite the system between
the two pseudo-spin states, restoring the $SU(2)$ invariance below
the PG dome. Conversely, an applied magnetic field breaks the $SU(2)$
symmetry in favor of the CDW state and beyond a certain critical field
$B_{c}$, a ``pseudo spin-flop'' is observed where the ground state
``flips'' from the SC state to CDW order. This ``pseudo spin-flop''
was precisely observed in experiments performed under magnetic field,
with a critical field $B_{c}\sim17T$ \cite{Doiron-Leyraud07,Sebastian12,Wu13}.
In particular, the ultra-sound experiment \cite{LeBoeuf13} shows that 
the typical \textbf{B} versus T phase diagram in Fig. \ref{fig:tb} is very
similar to Fig.\ref{fig:so5}-2). Within the EHS model, this experiment
was addressed in Ref. \cite{Einenkel14}. Note that a co-existence
phase is present in this phase diagram, which accentuates the similarity
with the phase diagram 2) in Fig. \ref{fig:so5} of the $SO(5)$ theory.
Notice as well that the CDW and SC temperatures are of the same order
of magnitude, which was never the case for the AF and SC states. It
is another indication that the $SU(2)$ symmetry is more likely verified in the underdoped cuprates than the $SO(5)$ symmetry.

Of course, a question can be raised at this point, which is that the
exact realization of the $SU(2)$ symmetry within the EHS model gives
a charge wave vector on the diagonal, while only axial charge order
was experimentally observed \cite{Hoffman02,Wise08,Fujita14,Ghiringhelli12,Shen05,He11,Achkar13,Blanco-Canosa14,Blackburn13a}.
It is an important question in the $SU(2)$ theory
and we will address it in details in the next section. For the moment
let us notice that similar rotations as in Eqn.(\ref{eq:1}) can be
generated for the axial wave vector $\mathbf{Q}=\{\mathbf{Q}_{x},\mathbf{Q}_{y}\}$
observed experimentally, which rotates similar multiplets as in Eqn.(\ref{eq:5})
but for the axial wave vector. This idea of a rotation between the
$d$-wave SC state and the axial charge order \cite{Hayward14} was used
to explain that the CDW signal is peaked at $T_{c}$ \cite{Ghiringhelli12,Blackburn13a}.
It was also used in explaining the $A_{1g}$ mode observed in Raman
scattering as a collective mode associated to this specific rotation
\cite{Montiel15a,Sacuto13,SacutoSidis02}.

The notion an emergent symmetry is more general than any of its specific
representations. It is indeed very nice to have a model, although
very simplified, where the $SU(2)$ symmetry is exactly realized (at
all energy scales), but the main concern is whether this symmetry
is approximately realized at finite temperatures in the underdoped
region of the phase diagram. That is the interest of the concept of
emergent symmetry: although it can be exactly realized in only a
few effective models, if the splitting between the two pseudo-spin states is
smaller than the typical energy of each state, it can also be approximately
realized at low energies in the more realistic 2D $t-t'$ Hubbard model (this was verified explicitly in Refs. \cite{deCarvalho:2014tj,Freire:2015kg} using two-loop RG techniques).

Another remark that can be made at this stage, is that another type
of $SU(2)$ symmetry was identified early on, which consists of performing
a particle-hole transformation on each site $c_{i\sigma}^{\dagger}\rightarrow c_{i-\sigma}$,
which translates in the reciprocal space as $c_{k\sigma}^{\dagger}\rightarrow c_{k-\sigma}$
for all k vector. This symmetry is interesting for the phase diagram
of the cuprates because it is exact at half-filling and will be gradually
broken with doping \cite{Kotliar88a,Lee06}. The operators for this
symmetry group take the form \begin{subequations} \label{eq:1-3}
\begin{align}
\eta_{ph}^{+} & =\sum_{\mathbf{k}}c_{\mathbf{k}\uparrow}^{\dagger}c_{\mathbf{k}\downarrow}^{\dagger}\\
\eta_{z} & =\sum_{\mathbf{k}\sigma}\left(c_{\mathbf{k}\sigma}^{\dagger}c_{\mathbf{k}\sigma}-1\right),
\end{align}
\end{subequations} while one irreducible $l=1$ representation associated
to it can be taken as \footnote{Note that a different representation was used in the slave-boson approaches
\cite{Kotliar88a,Lee06}, where the rotation was performed from the
$d$-wave SC state towards $\pi$-flux phases.} \begin{subequations} \label{eq:5-1} 
\begin{align}
\Delta_{1} & =-\frac{1}{\sqrt{2}}\sum_{\mathbf{k}}d_{\mathbf{k}}  c_{\mathbf{k}\uparrow}^{\dagger}c_{-\mathbf{k}\downarrow}^{\dagger}  ,\\
\Delta_{0} & =\frac{1}{2}\sum_{\mathbf{k},\sigma}d_{\mathbf{k}}  c_{\mathbf{k}\sigma}^{\dagger}c_{\mathbf{-k}\sigma}  ,\\
\Delta_{-1} & =\frac{1}{\sqrt{2}}\sum_{\mathbf{k}}d_{\mathbf{k}}  c_{\mathbf{k}\downarrow}c_{-\mathbf{k}\uparrow}  .
\end{align}
\end{subequations} Interestingly, within the EHS model, the operators
(\ref{eq:1-3}) and (\ref{eq:5-1}) are, respectively, the same as (\ref{eq:1})
and (\ref{eq:5}) with a diagonal wave vector, since there the summation
over \textbf{k} is reduced to the eight hot spots. The EHS model is also an exact realization of the $SU\left(2\right)$ symmetry associated
with particle-hole transformation.

Using (\ref{eq:1-3}) one can also ask oneself what is the $SU(2)$
partner of the observed axial CDW. To fix the ideas let us take a
uni-axial CDW order with ordering wave vector $\mathbf{Q}_{x}$ relating
two hot spots. One can then construct the $l=1$ irreducible representation
using the particle- hole transformation. This gives \begin{subequations}
\label{eq:5-1-1} 
\begin{align}
\Delta_{1} & =-\frac{1}{\sqrt{2}}\sum_{\mathbf{k}}d_{\mathbf{k}}  c_{\mathbf{k}\uparrow}^{\dagger}c_{-\mathbf{k+Q}_{y}\downarrow}^{\dagger}  ,\\
\Delta_{0} & =\frac{1}{2}\sum_{\mathbf{k},\sigma}d_{\mathbf{k}}  c_{\mathbf{k}\sigma}^{\dagger}c_{\mathbf{k+Q}_{x}\sigma}  ,\\
\Delta_{-1} & =\frac{1}{\sqrt{2}}\sum_{\mathbf{k}}d_{\mathbf{k}}  c_{-\mathbf{k+Q}_{y}\downarrow}c_{\mathbf{k}\uparrow}  ,
\end{align}
\end{subequations} which means that the $SU(2)$ partner of the $\mathbf{Q}_{x}$
CDW is the pair density wave (PDW), namely a non zero center of mass
SC state, with $\mathbf{Q}_{y}$ wave vector. This notion of PDW state
was introduced recently to explain the very unusual ARPES data tracing
the formation of the PG in Bi-2201\cite{Lee14,Agterberg:2014wf}.
In this theory, the formation of the PDW is suggested as the primary
mechanism for the formation of the PG state, which means that the
observed CDW is a secondary order. As such, it should be observed
at a wave vector twice as big as the PDW wave vector $\mathbf{Q}_{CDW}=2\mathbf{Q}_{PDW}$.
In contrast, if the mechanism governing the underdoped region is a
hidden pseudospin $SU(2)$ symmetry, then the partner of a bi-axial CDW is a
bi-axial PDW with the same wave vectors $\mathbf{Q}_{CDW}=\mathbf{Q}_{PDW}$
(see the Refs. \cite{Pepin14,Freire:2015kg,Wang15b,Carvalho15b}). The latter scenario has recently been verified experimentally
\cite{Hamidian16}.

\section{Non linear $\sigma$-model, and $SU(2)$ vs. $U(1)$ pairing fluctuations}
\label{sectionfluc}
The idea of emergent symmetries received a recent critique, that when
the group of symmetry is large enough, the symmetric phase is unstable
to smaller subgroups \cite{Fradkin15}. For example, the symmetric
phase associated to the $SO(5)$ symmetry which was intended to describe
the PG shall decompose into the $SU(2)\times U(1)$ group describing
fluctuations around the AF and SC phases respectively. Similarly,
the $SU(2)$ symmetry which rotates between the CDW and SC channel
shall decompose into the $U(1)\times U(1)$ groups. In this section,
we consider seriously the criticism that the symmetric phase of large
non-abelian groups is unstable, but wonder more particularly about
the fate of SC fluctuations.

The role of SC fluctuations in the physics of cuprates is indeed very
mysterious. We know that they are a few orders of magnitude more intense
that in standard metals like $Al$, or $Cu$ \cite{EmeryVJ:1995dr},
but experiments detecting pure the Josephson effect were observed
only a few tens of degrees above $T_{c}$ \cite{Bergeal:2008gf,Alloul:2010ko,RullierAlbenque:2011ji}.
In the deeply underdoped phase, $U(1)$ SC fluctuations form a dome
shape that we will discuss further in this section \cite{RullierAlbenque:2011ji,Li:2011cs,Li:2013ed,Wang:2002ke,Wang:2006fa}
. Direct observation of pre-formed pairs in the PG phase was always
negative, but a giant proximity effect was observed in the Lanthanum-compounds
induced in the PG phase when it is surrounded by optimal SC phases
\cite{Decca:2000hc,Bozovic:2004cp,Morenzoni11}. The very easy injection
of pairs from optimally doped into the PG phase suggests that the
PG phase is related to the SC phase through a hidden symmetry. Such
proximity effects were predicted in the case of the $SO(5)$ symmetry
and never observed \cite{Demler:1998iw}, but specific predictions
in the case of the $SU(2)$ symmetry were never discussed in detail.

\subsection{The $SU(2)$ SC fluctuations}

\begin{figure}[tb]
\begin{minipage}[c]{4cm}%
 a)\includegraphics[width=27mm]{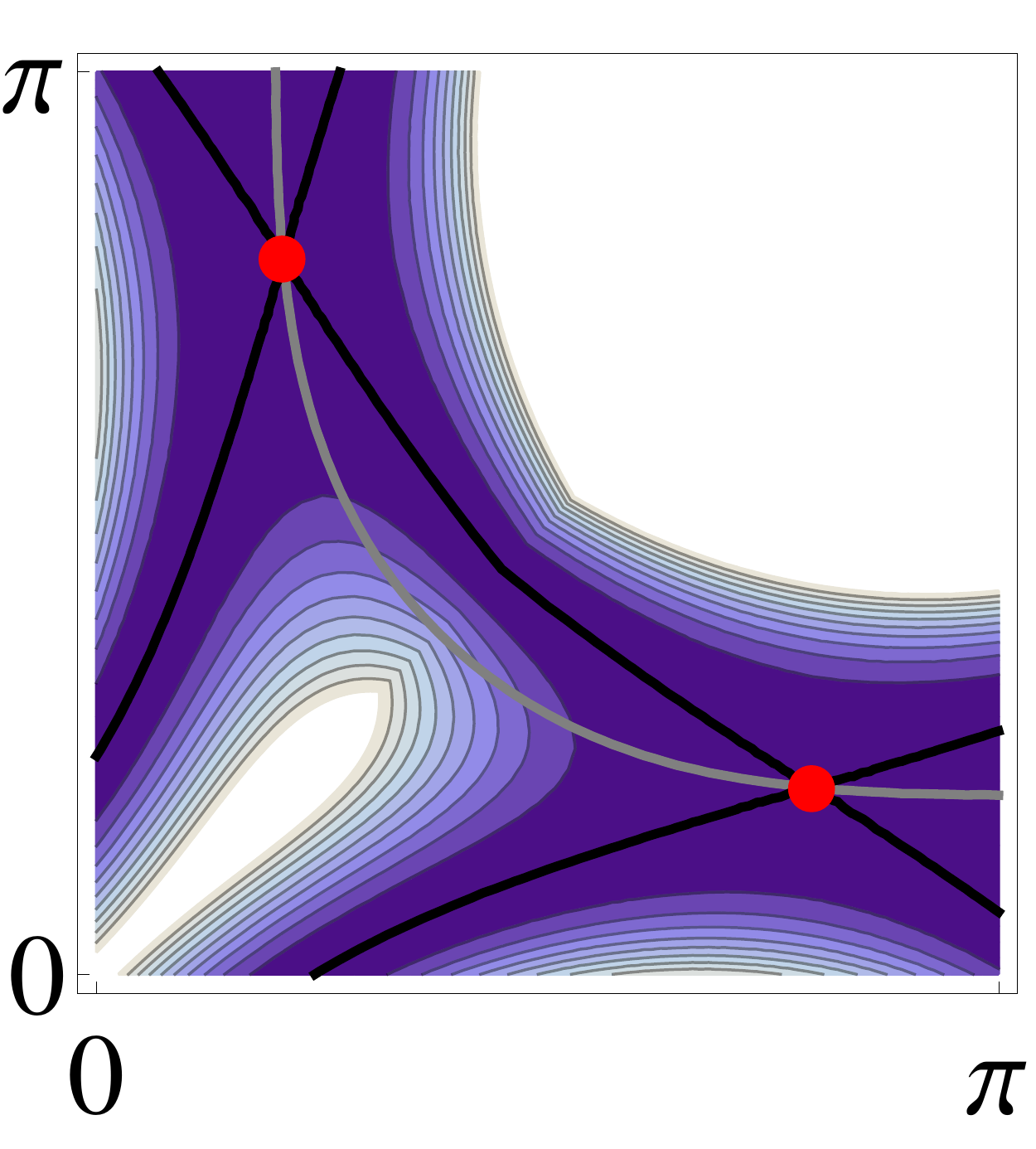} \vspace{1ex}
\end{minipage}%
\begin{minipage}[c]{4cm}%
 b)\includegraphics[width=40mm]{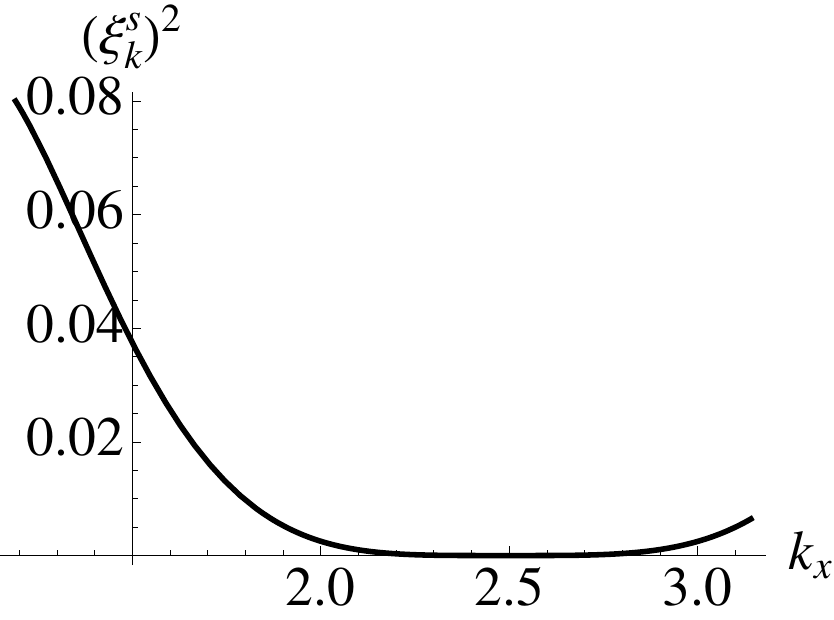} \vspace{1ex}
\end{minipage}\caption{\label{fig:symbreak} (Color online) a) Visualization of the SU(2)
symmetry breaking contribution $(\xi_{{\bf {k}}}^{s})^{2}$ in the
positive region of the first Brillouin zone. It is small in the blue
region and vanishes for the two black lines crossing the hotspot,
but grows in the nodal line to the upper edge. b) Variation of $(\xi_{{\bf {k}}}^{s})^{2}$
along the Fermi surface (shown as a gray line in panel a)) parametrized
as a function of $k_{x}$. While $(\xi_{{\bf {k}}}^{s})^{2}$ vanishes
at the hotspot and stays small at the antinodes at $k_{x}=\pi$, it
grows at the left side when approaching the nodal position.}
\end{figure}

In this section we assume that at an intermediate energy scale SC
fluctuations are present, protected by an $SU(2)$ symmetry between
the CDW and SC channel. The microscopic derivation of the non linear
$\sigma$-model describing the $SU(2)$ fluctuations can be found,
in the context of the EHS model in Ref.\cite{Efetov13}, and in the
context where regions of the Brillouin zone instead of points are
``hot'' -or hot anti-nodal regions-, in Ref.\ \cite{Kloss15a}.
The massless $O(4)$ effective free energy has the following form 
\begin{align}
F_{SU(2)} & =\frac{T^{2}}{2}\sum_{\varepsilon,\omega}\int_{\mathbf{k},\mathbf{q}}tr\delta\hat{u}_{-k,q}^{\dagger}\left[J_{0,k}\omega^{2}+J_{1,k}q^{2}\right]\delta\hat{u}_{k,q}\label{eq:nlsm2-1-1}
\end{align}
where $\hat{u}_{\mathbf{k},\mathbf{q}}=\left(\begin{array}{cc}
\Delta_{CDW} & \Delta_{SC}\\
-\Delta_{SC}^{*} & \Delta_{CDW}^{*}
\end{array}\right)$ is the $SU(2)$ matrix associated with the condition $\left|\Delta_{CDW}\right|^{2}+\left|\Delta_{SC}\right|^{2}=1$,
and the $tr$ runs on the $SU(2)$ structure. The coefficients write
$J_{0,k}=\left|M_{\mathbf{k}}\right|^{2}/\left|G^{-1}\right|^{2}$,
and $J_{1,k}=J_{0,k}v_{\mathbf{k}}^{2}$, where $M_{\mathbf{k}}$
is the magnitude of the mean-field $SU(2)$ order parameter $\hat{M}_{\mathbf{k},\mathbf{q}}=\left(\begin{array}{cc}
 & \hat{m}_{\mathbf{k,}q}\\
\hat{m}_{\mathbf{k,}q}^{\dagger}
\end{array}\right)_{\Lambda}$ , with $\hat{m}_{\mathbf{k},\mathbf{q}}=M_{\mathbf{k}}\hat{u}_{\mathbf{k},\mathbf{q}}$,
which has a $4\times4$ structure in the $\tau\times\Lambda$-$SU(2)$
spaces where $\tau$ is the particle-hole transformation and $\Lambda$
is the $\mathbf{Q}$-translation. The Green's function writes $\hat{G}^{-1}=\hat{G}_{0,k}^{-1}+\hat{M}_{k,0}$,
with $\hat{G}_{0,k}^{-1}=i\omega-\left(\tau_{3}\xi_{\mathbf{k}}^{s}-\xi_{\mathbf{k}}^{a}\right)\Lambda_{3}$,
with $\xi_{\mathbf{k}}^{s,a}=\left(\epsilon_{\mathbf{k}}\pm\epsilon_{-\mathbf{k}-\mathbf{Q}}\right)/2$,
and $\epsilon_{\mathbf{k}}$ being the electron dispersion. Note that
no information was given on the value of the $\mathbf{Q}$-wave vector
for the CDW sector. It corresponds in all generality to the $SU(2)$
operators (\ref{eq:1}) and (\ref{eq:1-1}). The exact $SU(2)$ symmetry
is verified when $\xi_{\mathbf{k}}^{s}=0$ which effectively kills
the $\tau_{3}$-term in the equation for $\hat{G}_{0,k}^{-1}$. $\xi_{\mathbf{k}}^{s}$
hence models the symmetry breaking term associated with this specific
wave vector and contributing to the free energy as 
\begin{equation}
F_{SB}=\frac{T^{2}}{2}\sum_{\varepsilon,\omega}\int_{\mathbf{k},\mathbf{q}}J_{3,k}\,tr\left[\delta\hat{u}_{-k,q}^{\dagger}\tau_{3}\delta\hat{u}_{k,q}\tau_{3}\right],\label{eq:10a}
\end{equation}
with 
\begin{align}
J_{3,k}= & \frac{1}{4}\frac{\left|m_{0,k}\right|^{2}\left(\xi_{\mathbf{k}}^{s}\right)^{2}}{\left|G^{-1}\right|^{2}}.\label{eq:10}
\end{align}
The shape of the symmetry-breaking term Eqn.(\ref{eq:10}) is visualized
in Fig.\ \ref{fig:symbreak}. One can observe the anisotropy of the
mass in various directions in the Brillouin Zone : the mass is much
bigger in the nodal direction than in the anti-nodal one. 

\subsection{Resonant Peierls Excitonic (RPE) state}

We now integrate the $SU(2)$ SC fluctuations out of the partition
function, and evaluate the consequences of them in the charge channel.
We get the following effective action 
\begin{equation}
S_{\text{eff}}[c]=\sum_{\mathbf{kk'q}}\pi_{k,k',q}c_{\uparrow\mathbf{k}}^{\dagger}c_{\uparrow\mathbf{k'}}c_{\downarrow-\mathbf{k+}q}^{\dagger}c_{\downarrow-\mathbf{k'}+\mathbf{q}},\label{eq:sfinfull}
\end{equation}
with 
\begin{equation}
\pi_{k,k',q}=\langle\bar{\Delta}_{k,q}\Delta_{k',q}\rangle=\frac{\pi_{0}\left(\delta_{k,-k'}+\delta_{k,k'}\right)}{\left(J_{0}\omega_{n}^{2}+J_{1}({\bf {v}}\cdot{\bf {q}})^{2}+a_{0,k}\right)},
\end{equation}
where the form of the SC fluctuations comes from Eqns.(\ref{eq:nlsm2-1-1},\ref{eq:10a}).
The self-consistent Dyson equation ( or ``gap equation'') writes
\begin{equation}
\chi_{k,k'}=\sum_{q}\pi_{k,k',q}[\hat{G}(q-k,q-k')]_{12},\label{eq:chieq3}
\end{equation}
with $\chi_{k,k'}=\sum_{q}\pi_{k,k',q}\langle c_{\sigma\mathbf{k}}^{\dagger}c_{\sigma\mathbf{k'}}\rangle$
and with 
\begin{align}
[\hat{G}_{k,k'}]_{12} & =-\langle c_{\sigma}^{\dagger}(k)c_{\sigma}(k')\rangle\nonumber \\
 & =-\frac{\chi_{k,k'}}{(i\epsilon_{n}-\xi_{{\bf {k}}})(i\epsilon_{n}'-\xi_{k'})-\chi_{k,k'}^{2}}.\label{eq:chieq2}
\end{align}
\begin{figure}[tb]
\begin{minipage}[c]{4cm}%
 a) \includegraphics[width=40mm]{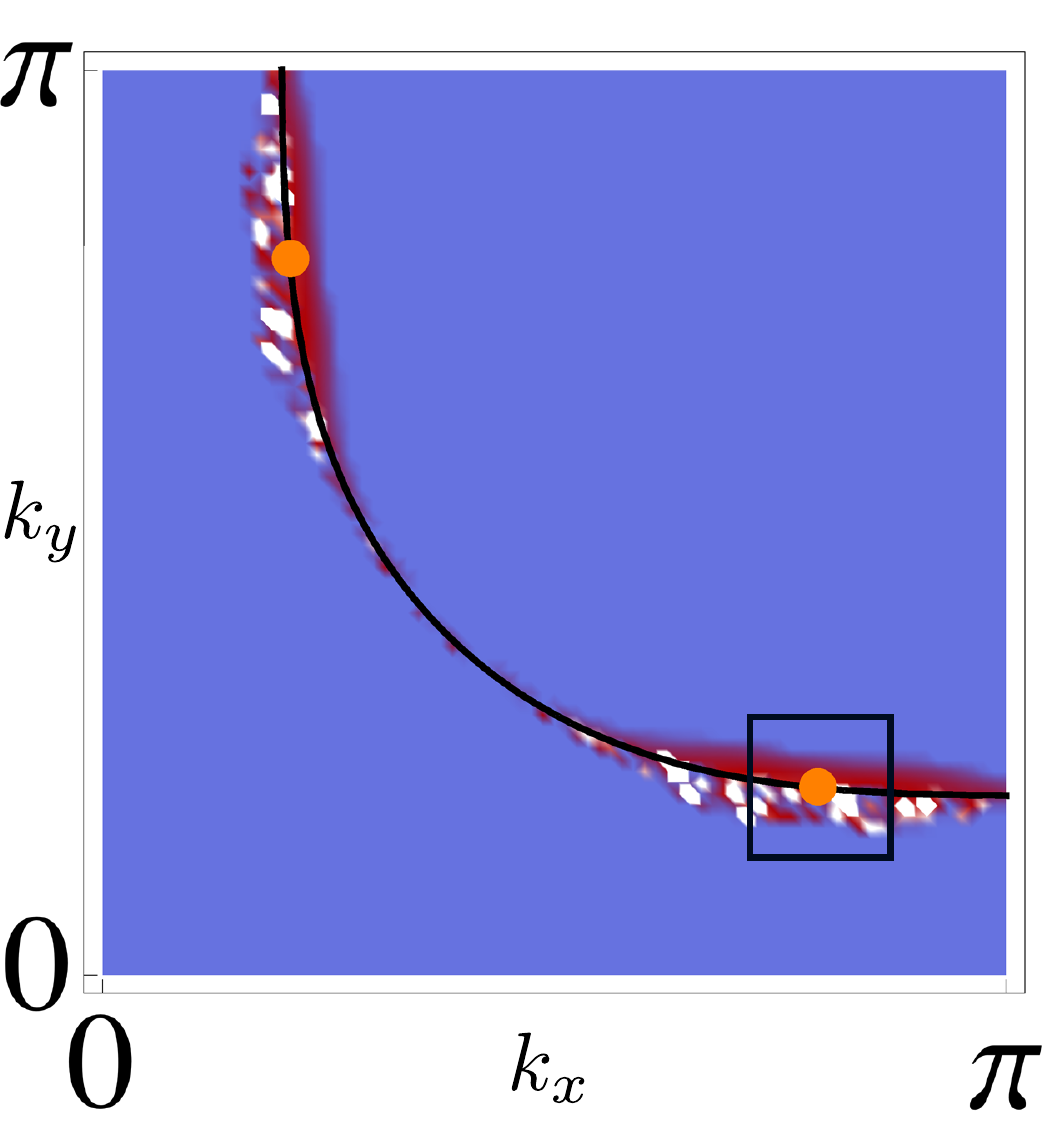} \vspace{-1ex}
\end{minipage}%
\begin{minipage}[c]{4cm}%
 b) \includegraphics[width=30mm]{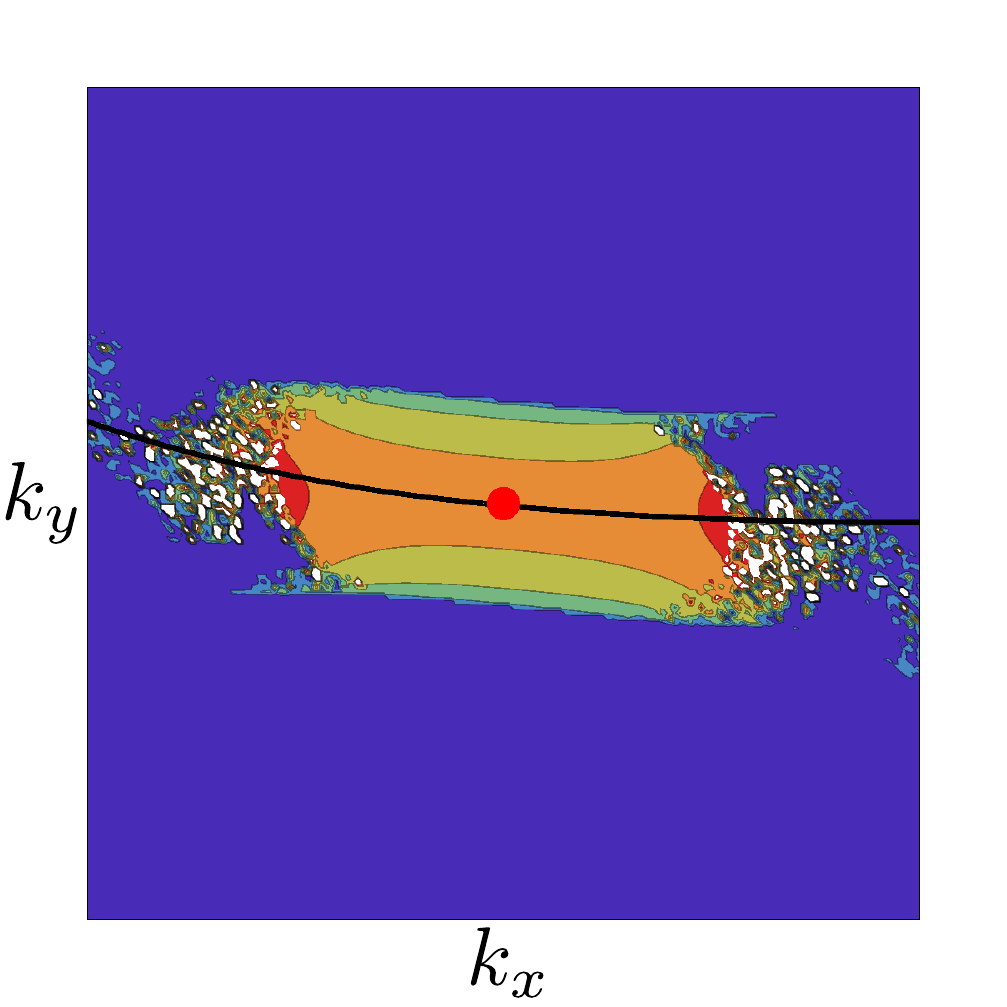} \vspace{1ex}
\end{minipage}\caption{\label{fig:1bz} (Color online) left) Density of the charge order
parameter $|\chi_{k,-k}|$ in the first Brillouin zone from the RPE
state. The charge density follows the Fermi surface, but due to a
SU(2) dependent mass contribution, does not stabilize in the nodal
region. right) Charge order parameter around the hotspot position
for a constant $\pf$ ordering vector. From Ref.\ \cite{Kloss15a}.}
\end{figure}
\begin{figure}[tb]
\includegraphics[width=26mm]{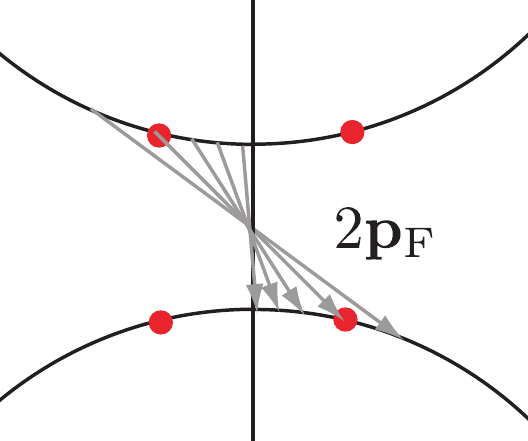} \vspace{-0.5ex}
 \caption{\label{fig:2pf} (Color online) Set of degenerate $\pf$ couplings
between electrons on opposed Fermi surfaces in the antinodal region
due to the RPE state. From Ref.\ \cite{Kloss15a}.}
\end{figure}

%\textcolor{blue}{
%It is instructive to visualize in real space what the RPE state looks like. For this purpose we evaluate the charge on the oxygen, which in real space is given by 
%\begin{equation}
%\rho({\bf {r}},a_{0}^{i})\sim\sum_{{\bf {k}},{\bf {P}},\sigma}\cos((k_{i}+P_{i})a_{0}^{i})\cos({\bf {r}\cdot{\bf {P}}})\langle c_{{\bf {k}+{\bf {P},\sigma}}}^{\dagger}c_{{\bf {k},\sigma}}\rangle,
%\end{equation}
%where $i\in\{x,y\}$. The oxygen positions around the copper are as shown in Fig.\ \ref{fig:cu_o_positions}. The excitonic state has an almost pure $d$-wave structure on the oxygen atoms, and axial modulation of them is seen with wave vectors very close to those observed experimentally.}

To get some idea about the nonlocal nature the order parameter $\chi_{\bf{k,k'}}$,
it is illustrative to consider one of the labels having a constant shift $\bf{k'}=\bf{k+P}$.
The order parameter can then be decomposed as $\chi_{\bf{P,k}} = \chi_{\bf{P}} F_{\bf{P}}$ \cite{Kloss15a},
where $\chi_{\bf{P}}$ is a plain wave and $F_{\bf{k}}$ a formfactor for the electron-hole pair with a finite extent, see Fig.\ \ref{fig:1bz}b).
The total order $\chi_{\bf{k,k'}}$ is thus a superposition of all local $\chi_{\bf{P,k}}$ orders,
where $\bf{P}$ is running over the whole set of degenerate $\pf$ couplings as visualized in Fig.\ \ref{fig:2pf}.
The gap in the antinodal region is $\Delta^{PG}_{\bf{k}} = \sum_{\{ \bf{P}\}} \chi_{\bf{P},k}$.
Note that in Eqns. (\ref{eq:chieq3},\ref{eq:chieq2}) the external
wave vectors $\mathbf{k},\mathbf{k'}$ are \textit{a priori} not defined, but
are let free to find self-consistently the most favorable solution.
We studied numerically the possible excitonic solutions of the gap
equations and the result is depicted in Fig.\ \ref{fig:1bz}. We
obtained an excitonic state in which a large number of wave vectors
are degenerate with a typically $\mathbf{k}-\mathbf{k'}=2\mathbf{k}_{F}$
which are spread out in the anti-nodal region of the Brillouin zone producing
a depletion of the density of states in this region (see Fig.\ \ref{fig:2pf}).
Due to the angular dependence of the fluctuation mass ($a_{0,node}\gg a_{0,anti-node}$),
we obtain a preferential gapping out of the anti-nodal region, which
is characteristic of the $SU(2)$ SC fluctuations compared to the
original $U(1)$. 

\subsection{Long range charge order}

At this stage we have proposed a theory for the PG phase of cuprates
superconductors. In the following we will give some more arguments
that this theory is a promising candidate to describe high T$_{c}$
superconductors. We know from a body of experimental evidence, though,
that the PG phase is distinct from the observed uni-axial CDW. At
zero field, the CDW dome decreases when the oxygen doping decreases,
which is at variance with the PG $T^{*}$-line, which increases with
the chemical potential (or oxygen doping) \cite{Achkar12,Achkar13,Bakr13,Ghiringhelli12,CyrChoiniere:2015wu,Blanco-Canosa14}.
Moreover, recent studies of the Fermi surface
reconstruction under a magnetic field of $17$\,T infer that the
PG is formed \textit{before} the Fermi surface is reconstructed by
CDW order -recall that the CDW becomes \textcolor{black}{long} range
and three dimensional beyond $B=17$\,T. There are many proposals
for the PG phase \cite{LeHur:2009iw,Rice12,Lee14,Fradkin15,Senthil03,Wang15b,Hayward14,Bulut2015,Nie2015,Gull:2013hh,Sordi2012,Sordi2012B},
but since ours consists of a special type of excitonic liquid, it
is important to shed light on the relationship between the excitonic
RPE phase and the observed axial CDW which is stabilized under magnetic
field.

Within the EHS
model, or within all sorts of simple weak coupling RPA evaluation,
we find that the axial charge order is a secondary instability, weaker
than CDW with a wave vector on the diagonal. The question that is
then raised, is why nothing at all is observed on the diagonal, whether
it is by STM \cite{Wise08,He14,Fujita14,Hamidian15a} or by X-rays
\cite{Blackburn13a,Ghiringhelli12,Achkar12,Blanco-Canosa14} The simplest
explanation is that the pseudogap is forming, gapping out the anti-nodal
region of the Brillouin Zone, and wiping out the CDW instability on
the diagonal. When the mechanism of formation of the pseudogap instability
has operated, then the secondary instability can be visible, at the
tip of the Fermi arcs. Many suggestions have been made for the formation
of axial CDW order. The fact that this wave vector is present as a
secondary instability in any weak coupling theory, and stabilized
for example, in the presence of additional effect like Coulomb interactions
\cite{Pepin14,Wang14,Allais14c}, within both one-loop and two-loop RG \cite{deCarvalho2013738,deCarvalho:2014tj,Whitsitt:2014eq,Carvalho15,Freire:2015kg}
or starting from a three band model \cite{Bulut2015}, or invoking
the proximity to the van Hove singularity \cite{Volkov:2015vh}
has been outlined in many works, including ours. All these studies
are based on the observation that axial CDW is distinct from the formation
of the PG state and starts to get formed at the tip of the arcs. In
all mentioned scenarios though, it is quit unclear why the CDW dome
is increasing with doping, in contrast with the PG line $T^{*}$.
The $SU(2)$-paradigm which related the axial CDW to SC fluctuations that  gives a good explanation for the maximum of amplitude of the
CDW order at $T_{c}$ \cite{Hayward14} seems to have been lost in
the attempts to rotate the leading wave vector from the diagonal to
the axes.
\begin{figure}[tb]
\includegraphics[width=50mm]{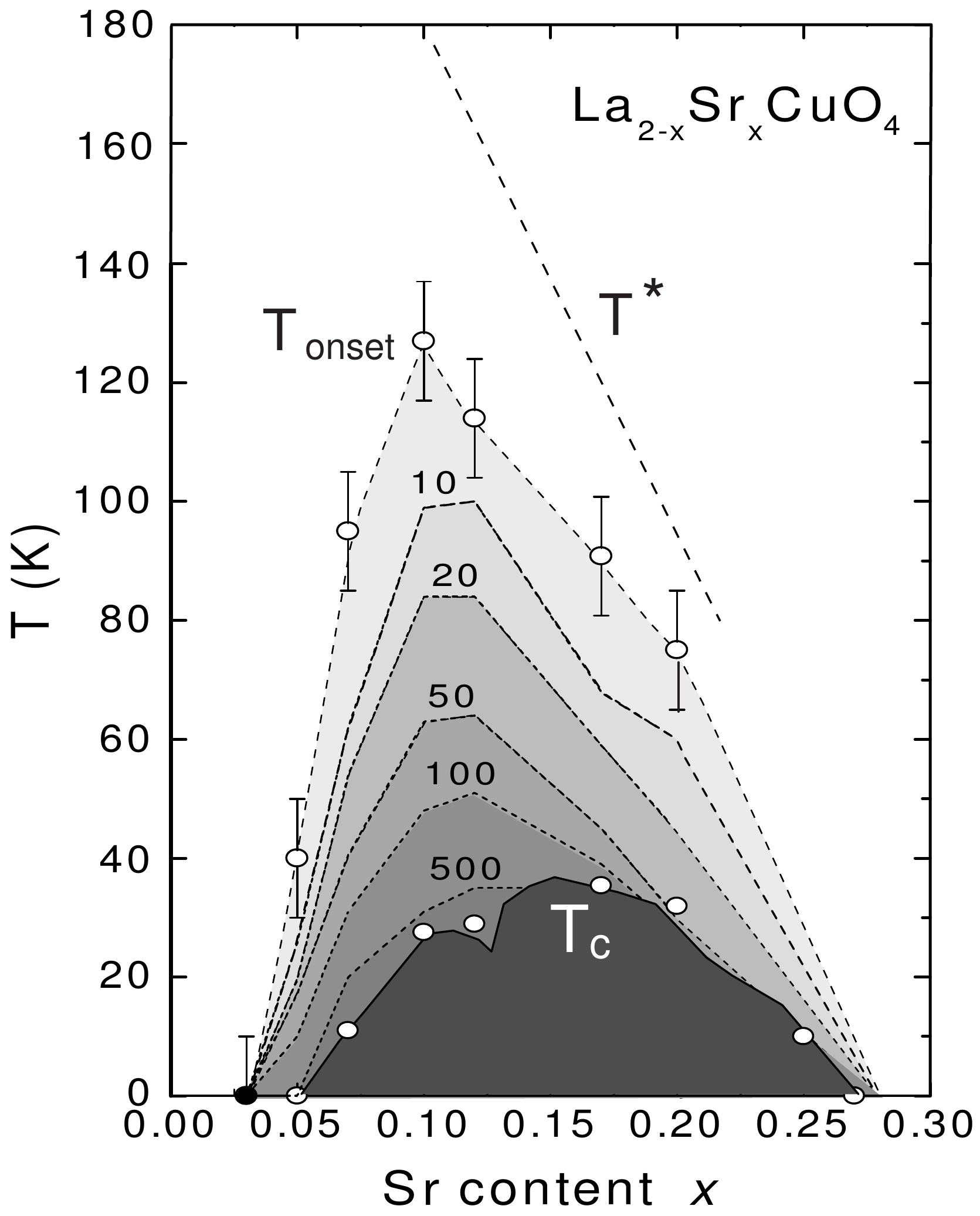} \vspace{-1ex}
\includegraphics[width=65mm]{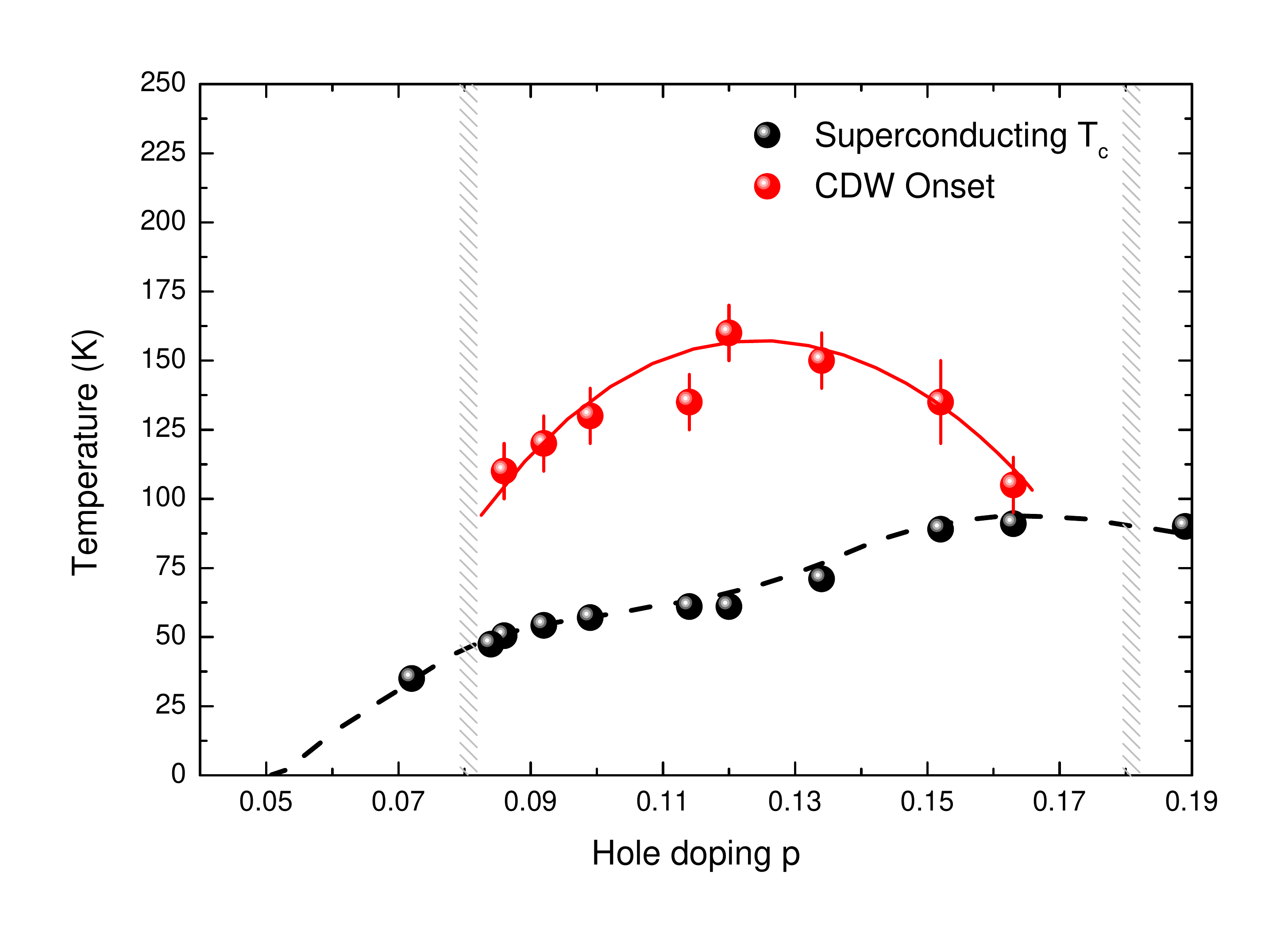} \vspace{-1ex}
\caption{\label{fig:nernst} (Color online) 
upper panel) Temperature-hole doping phase diagram deduced from Nernst effect experiments (from Ref.\cite{Wang:2006fa}). 
The gray area corresponds to CDW phase where Nernst coefficient is not zero.  
lower panel) Temperature-hole doping phase diagram with the superconducting critical temperature (black dots) and the onset temperature of the CDW axial order (red dots) deduced from RXS experiments in YBa$_{2}$Cu$_{3}$O$_{6+x}$ (extrated from Ref. \cite{Blanco-Canosa14}).}
\end{figure}
\begin{figure}[tb]
\includegraphics[width=40mm]{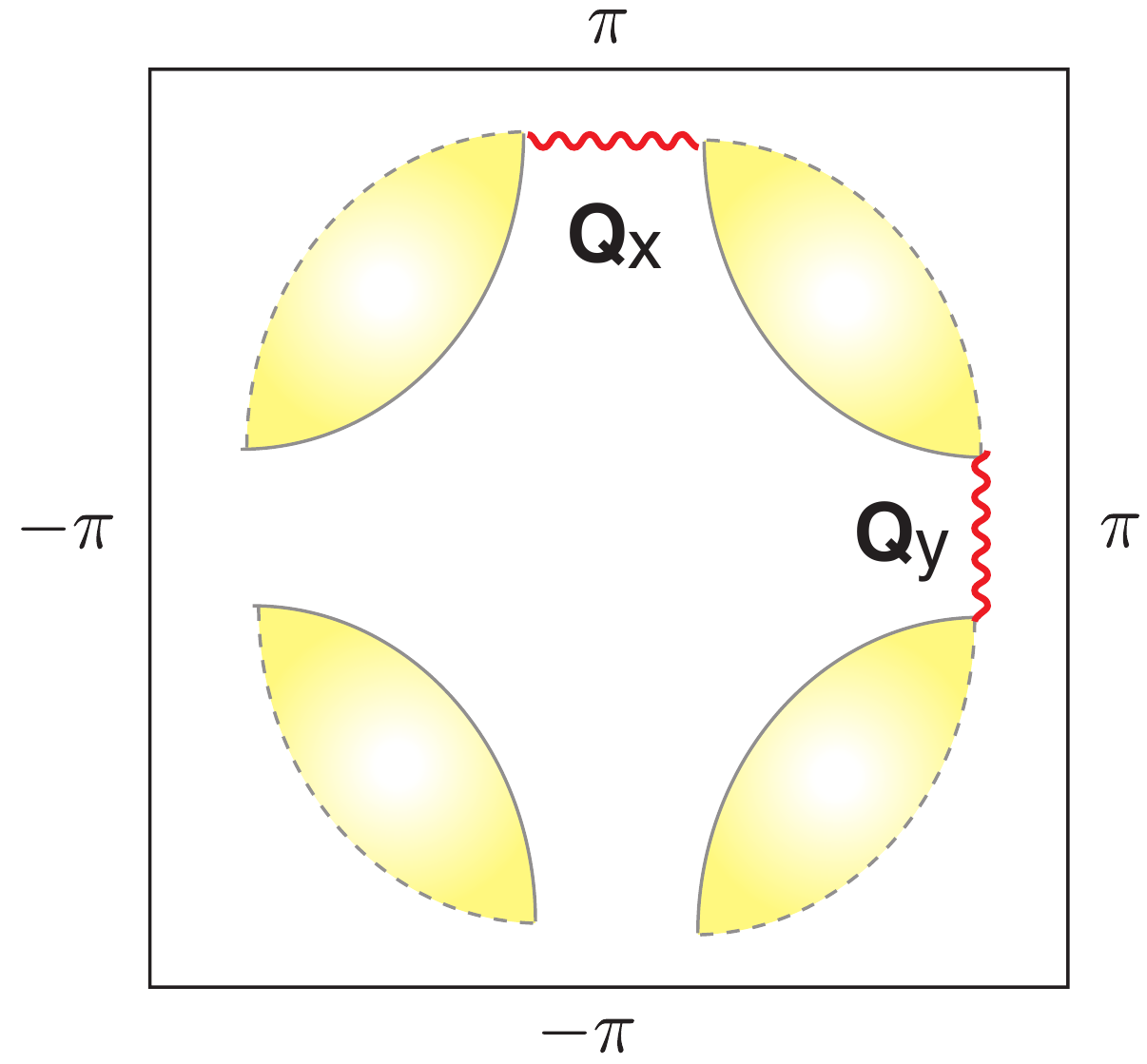} \vspace{-0.5ex}
 \caption{\label{fig:leggett} (Color online) Leggett mode (highlighted in red),
where charge order is induced by $U(1)$ phase fluctuations at the
tip of the arcs with the axial wave vector ${\bf{Q}}_{x/y}$.}
\end{figure}

In this paper, we would like to offer an alternative scenario which
to our knowledge has not been proposed yet. It is based on the observation
that the CDW dome follows closely the $U(1)$ SC fluctuation dome
that can be detected through many probes, like resistivity study \cite{RullierAlbenque:2000fl,Alloul:2010ko},
Nernst effect \cite{Wang:2002ke,Wang:2006fa,Li:2011cs}, Josephson
tunneling. The $U(1)$ standard SC fluctuations are usually very weak
in SC states because Coulomb interactions push them above the plasma
frequency\cite{Anderson:1963vi}. In the case of the optimally doped
cuprates, due to low dimensionality, phase fluctuations are observed
in a window of roughly 15\,K above $T_{c}$ \cite{Bergeal:2008gf}.
For more underdoped compounds, the approach of the Mott insulating
states produces a drastic increase of phase fluctuations due to the
phase-particle number Heisenberg uncertainty relations \cite{Emery95}.
We would like to exploit this idea here and suggest that the axial
CDW order is an effect of the $U(1)$ SC fluctuations. We use a recent
remark of Ref. \cite{Liu:2015uu} where it is noted
that after the formation of the PG, the Fermi surface has split into
four sections-or arcs, which are now independent from each other.
In these conditions, one can get a Leggett mode between the tips of
the arcs, which in turn can induce a charge order with the correct
wave vectors, see Fig.\ \ref{fig:nernst}. The gap equation
for the axial CDW mediated by the Leggett mode writes 
\begin{align}
\chi_{\mathbf{k}\sigma,\mathbf{k}+\mathbf{Q_{0}}\sigma} & =T\sum_{\omega_{n}\mathbf{q}}\pi_{\mathbf{Q_{0}},\mathbf{q}}[G_{-\mathbf{k}+\mathbf{q},-\mathbf{k}-\mathbf{Q_{0}}+\mathbf{q}}]_{12},\label{eq:15}
\end{align}
with $\mathbf{Q_{0}}=\mathbf{Q_{x/y}}$ being the axial wave vector,
$G_{12}$ given by Eqn.(\ref{eq:chieq2}) with the replacement $\xi_{\bf{k}} \rightarrow \xi_{\bf{k}} + \Delta^{PG}_{\bf{k}}$ to take the gapping of the FS into account, whereas $\pi_{\mathbf{Q_{0}},\mathbf{q}}$
is the $U(1)$ correlation of the phase fluctuations at the tip of
the arcs, as represented in Fig.\ \ref{fig:leggett} 
\begin{align}
\pi_{\mathbf{Q_{0}},\mathbf{q}} & =\left\langle {\cal T}\Delta_{\mathbf{k},\mathbf{-k+q}}\Delta_{\mathbf{k+Q_{0},-k-Q_{0+q}}}^{\dagger}\right\rangle _{U(1)} \nonumber \\
& = \frac{\pi_0}{J_0' \omega_n^2 + J_1' ({\bf{v}} \cdot {\bf{q}})^2 + m_0} ,
\end{align}
with $\mathbf{k}$ being the wave vector at the tip of the one Fermi
arc and $\mathbf{k}+\mathbf{Q_{0}}$ being the wave vector at the
tip of the adjacent Fermi arc. We use a generic form of the propagator $\pi_{\mathbf{Q_{0}},\mathbf{q}}$,
where $\pi_0, J_0', J_1'$ and $m_0$ are mon-universal parameters the dependence on ${\bf{Q}}_0$ is neglected.  We have performed a numerical study
of Eqn.(\ref{eq:15}) which confirms that $U(1)$ SC fluctuations
mediated by a Leggett mode produce axial CDW with the desired wave
vector. This proposal has the merit to consistently link both the
formation of the PG and the observed axial CDW to SC fluctuations,
the former being described by the $SU(2)$ non linear $\sigma$-model
while the latter are the standard $U(1)$ phase fluctuations.

\section{Discussion}
\label{sectionDiscussion}
In the phase diagram of high temperature cuprates a few key players
can be identified \cite{Norman03,Lee06}. There is at half-filling
the Mott insulating transition with typical energy of 1\,eV associated
to it. Antiferromagnetism is ubiquitous in the whole phase diagram,
with an ordered phase of typically $T_{Neel}\approx 700\,K$ at half-filling, very close
to the Mott transition, and strong, but short range AF fluctuations
in the underdoped regime. In the proposal of this paper, the mysterious
PG phase of high temperature cuprates is attributed to a new kind
of excitonic state, the RPE, which can be understood as a new type
of ``liquid'' of excitons, with a superposition of degenerate wave
vectors. This state is a consequence of integrating out the SC fluctuation,
protected by an emergent $SU(2)$ symmetry between the SC and charge
channel. In the discussion of this proposal, the first thing to recall
is that although antiferromagnetism is not directly responsible for
the PG, it is nevertheless the underlying force driving the emergence
of precursor orders. In the early version of this theory, the EHS
model has been studied as a reference model where the $SU(2)$ symmetry
is verified \cite{Metlitski10b,Efetov13}. In this model the eight
hot spots are singled out of the Fermi surface, and long range AFM
fluctuations stabilize the composite $SU(2)$- order parameter, composed
by a diagonal quadrupolar density wave and SC. In more generic versions
of this theory, the model is extended to ``hot regions'' of the
Brillouin zone - the anti-nodal regions, where AF acts predominantly
and the $SU(2)$ symmetry is most strongly verified \cite{Kloss15}. Antiferromagnetism
did not disappear from the phase diagram, but rather has a very special
relation to the PG by defining the width of the ``hot regions'',
thus limiting the domain of action of the RPE state, and also being
the driving force both behind SC pairing and the $SU(2)$ symmetry.

The concept of emergent symmetry though, is more robust and general
than even the idea of Quantum Criticality and it is under such a generic
paradigm that we want to cast out the underdoped region of cuprate
superconductivity. The main idea is that charge orders are the natural
partner and competitors of SC pairing in the underdoped region of
the cuprates, and typical pseudo ``spin flops'' between the two
orders are to be expected, and we believe already observed under magnetic
field \cite{LeBoeuf13}.

The experimental consequences of a phase diagram controlled by an
$SU(2)$ emergent pseudospin symmetry are numerous, and it is very likely that our proposal
for the RPE state may be confirmed or in-firmed within the next few years.

\subsubsection{Spectroscopic signatures}

\begin{figure}[hb]
a)\includegraphics[width=50mm]{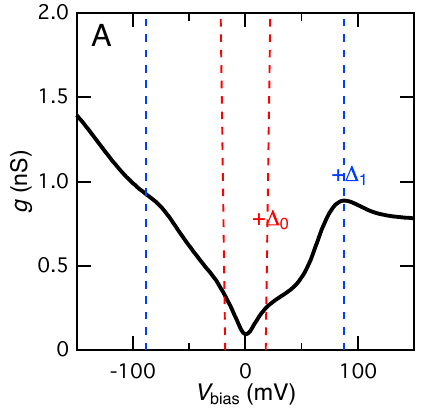} \vspace{1ex}
b)\includegraphics[width=50mm]{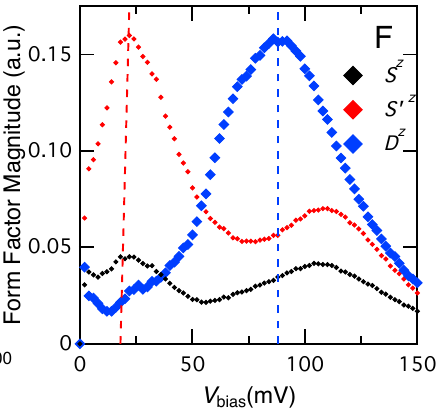} \vspace{1ex}
\caption{\label{fig:d-temp} (Color online) Upper panel a) Tunneling conductance measurements
from Ref.\ \cite{Hamidian15a} of underdoped cuprates. Two characteristic energies, a lower one for Bogoliubov quasiparticles
and a higher one corresponding to the pseudogap are observed.
Lower panel b) Energy dependence of the $s$- and $d$- wave form factors, indicating
that the higher gap-energy scale corresponds to the $d$-wave form.
From Ref.\ \cite{Hamidian15a}}
\end{figure}

One can first ask about the spectroscopic signature of such an excitonic
state.
What can be seen in STM or X-rays. Our claim here is that we
can reproduce the very recent findings on Bi-2212 \cite{Hamidian15,Davis16},
that the pure $d$-wave component of the axial CDW extends up to the
PG temperature, see Fig.\ \ref{fig:d-temp}. 
In the RPE state, indeed, the excitons form not only
around many degenerate wave vectors, but with a finite width around
each wave vector. 
The real space picture is that the particle-hole
pairing is non local in space, and modulated by many wave vectors.
When the induced charge on the oxygen is evaluated and Fourier transformed,
one finds that it is 90\% $d$-wave (100 \% for the diagonal wave vector
and a bit less for the others), and at the same time, the axial wave
vectors are more favored compared to the diagonal due to its nesting
properties in the anti-nodal region \cite{Montiel16}. The consequence
is that the charge on the Oxygen shows a preponderant spectrum with
axial wave vectors $\mathbf{Q}_{x}$ and $\mathbf{Q}_{y}$. At this
stage our conclusion is that the RPE state is already observed by
STM and X-rays, which have captured its preponderant contribution
on the axial wave vectors.

\subsubsection{Proximity effect}

\begin{figure}[h]
\includegraphics[width=30mm]{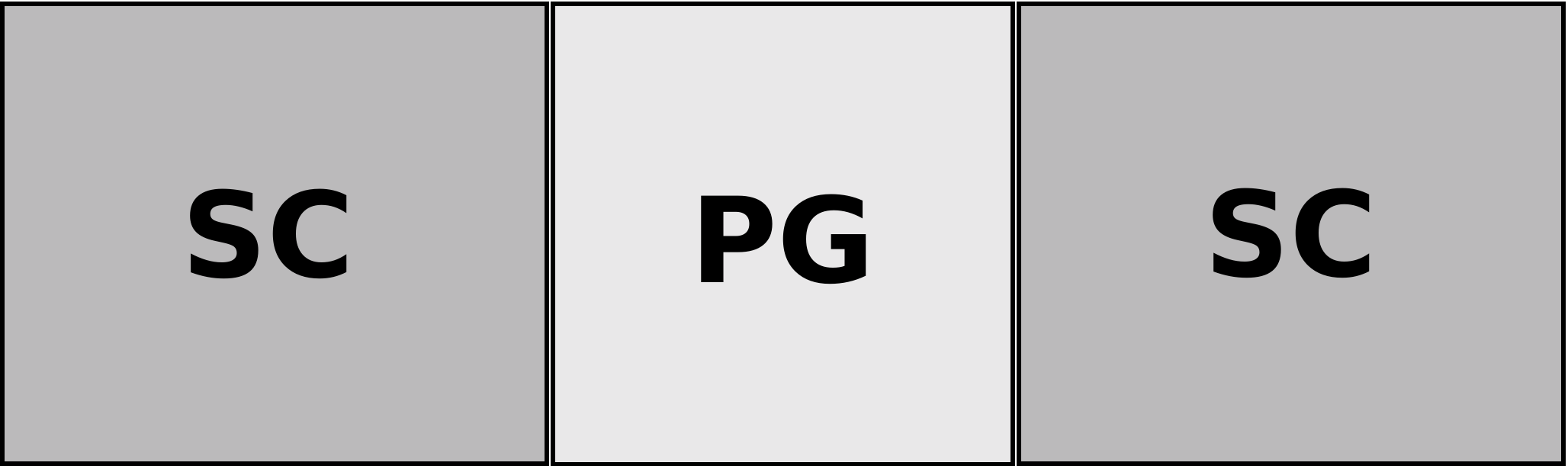} \vspace{-0.5ex}
 \caption{\label{fig:junction} (Color online) Proposition of a SC-PG-SC junction
to study the giant proximity effect within the SU(2) theory. At a
given temperature $T$ that is homogeneous over the junction, the
two outer SC layers are superconducing and at optimal doping such
that $T<T_{c}$. The inner PG layer is an underdoped SC in the pseudogap
phase ($T^{*}>T>T_{c}$). From the giant proximity effect we expect
the PG phase to become superconducting by lowering the thickness of
the inner layer. }
\end{figure}

A second remark is that emergent symmetries rotating
the SC phase to another type of order predict proximity effects when
the PG phase is sandwiched between two optimally doped superconductors.
The intensity of the induced current in the junction persists for
thickness of the gap material much greater than the superconducting
correlation length. This ''Giant'' Proximity Effect (GPE) is not
explicable by the standard theory of the proximity effect between
two SC junction, but can be understood in the situation where the
SC state is ''quasi-degenerate'' to another phase of matter and Cooper
pair can thus be easily injected from the SC state to the other state.
The situation is thus very promising for emergent symmetries, and
has been extensively studied in the case of the $SO(5)$-symmetry
\cite{Demler04,denHertog:1999dk}, where specific predictions for
the current as a function of the phase difference across the junction
can be made as well for the $SU(2)$ symmetry, see Fig.\ \ref{fig:junction}.
Note that a giant proximity effect has already been observed in various
compounds but has not been observed for the specific setup of the
$SO(5)$-group. One straightforward application of our theory is to
check whether $SU(2)$ fluctuations, where the rotation is between
the SC state and the CDW state can account for the experimental data
\cite{Tarutani91,Yuasa91,Kasai92,Meltzow97,Decca00,Bozovic04,Morenzoni11}.

\subsubsection{Magnetic field phase diagram}

\begin{figure}[h]
\includegraphics[width=65mm]{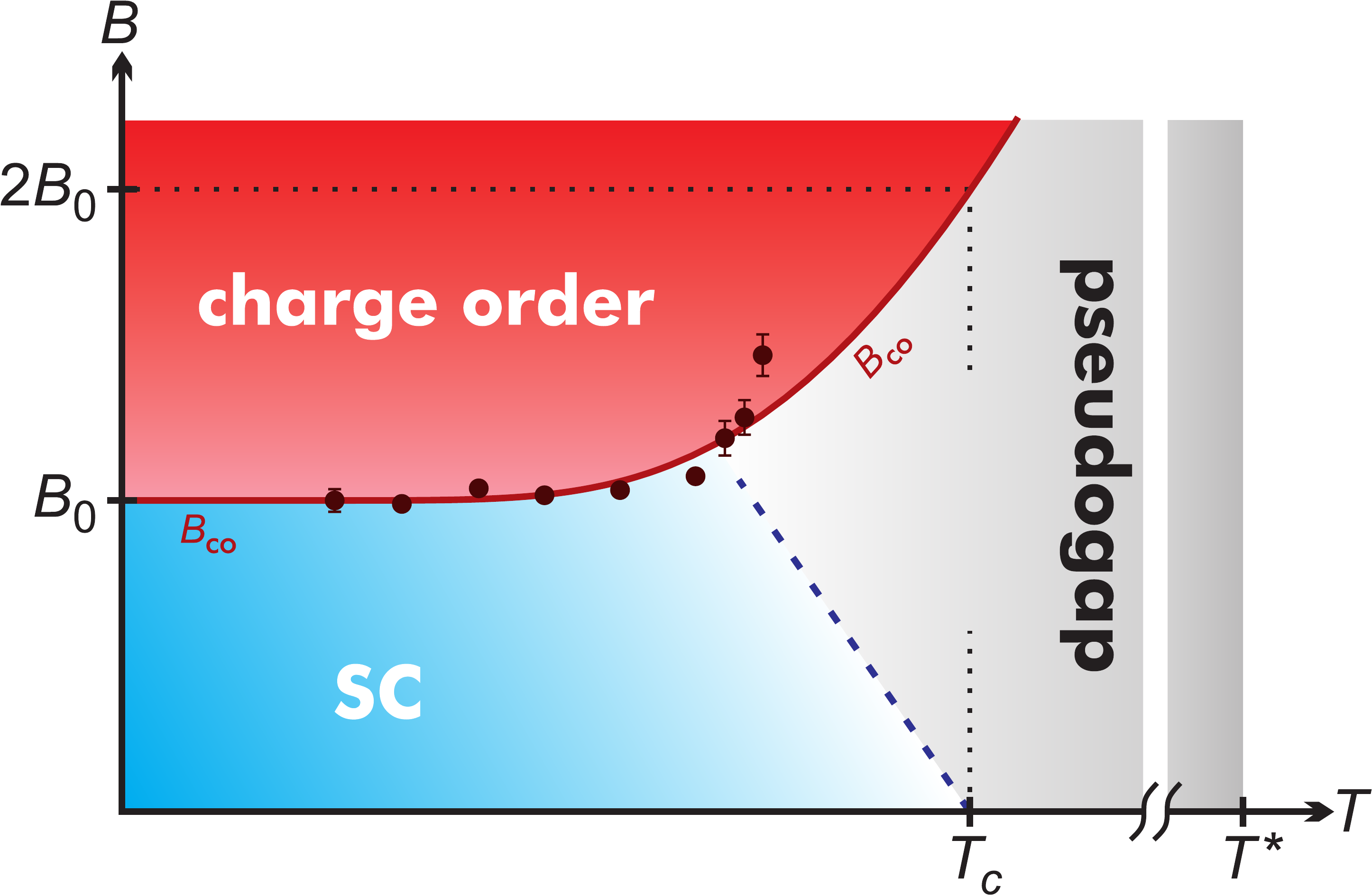} \vspace{-1ex}
\caption{\label{fig:tb_theo} (Color online) $B$-$T$ phase
diagram obtained from the spin-fermion model considering order parameter
fluctuations around the mean-field value with a nonlinear $\sigma$
model from Ref.\ \cite{Meier13}.}
\end{figure}

The phase diagram found as a function of magnetic field and temperature,
derived with a variety of experiments \cite{LeBoeuf13,Doiron-Leyraud07,Sebastian12,Wu13a,Wu:2015bt}
is typical for a super ``spin-flop'' between two states related
by a symmetry (see Fig.\ref{fig:tb_theo}). Note that three dimensional CDW has been recently observed
by X-ray scattering above $B=17$T \cite{Chang16}. 
The CDW and SC
orders have the same order of magnitude in this diagram, and the transition
between the two is very sudden, like in a generic spin-flop XY model
\cite{Meier13,Einenkel14}. Moreover, an $SU(2)$ partner of the axial
CDW has recently been reported, i.e.\ the PDW with the same wave vector
\cite{Hamidian16}. Although it is not a direct proof of the underlying
symmetry, it seems to rule out other scenarios for the PG state where
the PDW is primary while the CDW orders are secondary, and hence occur
at twice the same wave vector as the PDW.

\subsubsection{Collective modes}

\begin{figure}[t]
\includegraphics[width=60mm]{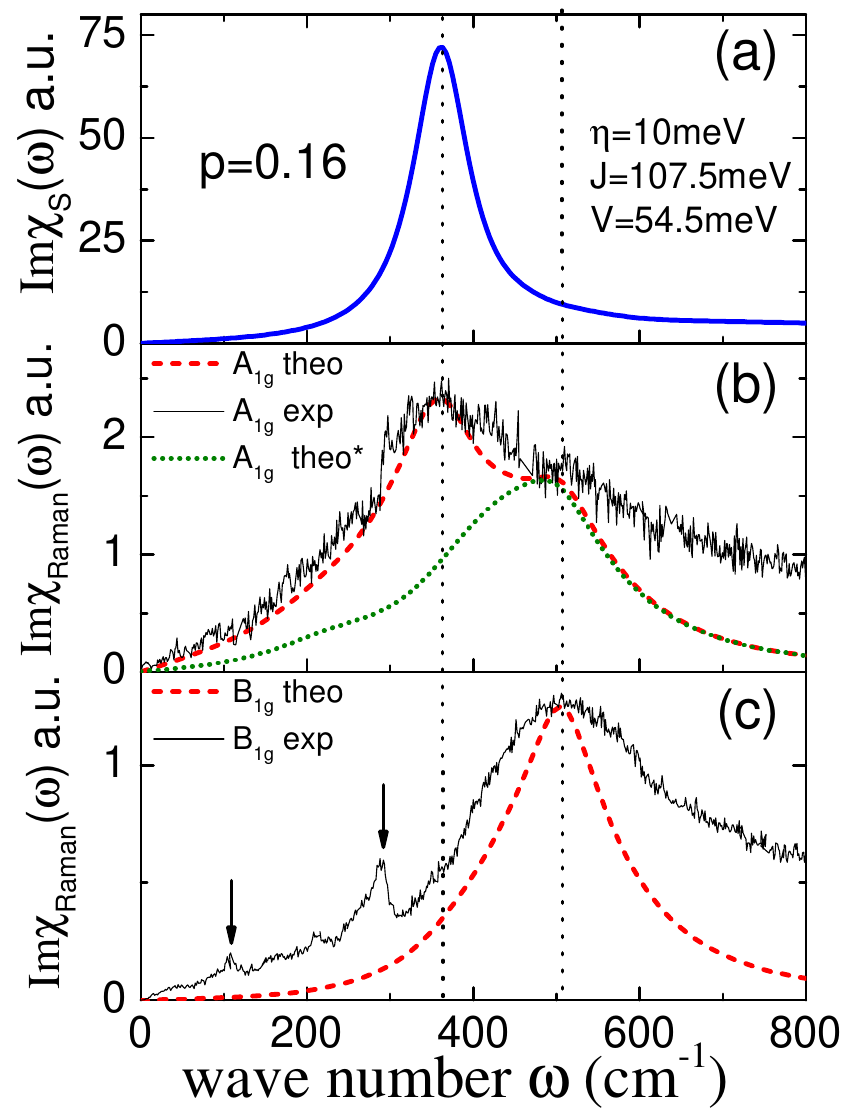} \vspace{-0.5ex}
 \caption{\label{fig:raman} (Color online) 
 Raman scattering response from a collective mode.
 a) shows the calculated Neutron susceptibilities at the momentum ${\bf{Q}}=(\pi,\pi)$ as a function of the frequency. 
 In b) (and c) ), the experimental (solid line) and calculated (dashed line) Raman response in the A$_{1g}$ ($B_{1g}$) Raman channels. 
 The Raman resonance arises at the same frequency than the Neutron resonance at ${\bf{Q}}$  (Fig (a) and (b)) 
 below the superconducting coherent peak energy observable in the $B_{1g}$ symmetry (Fig (b) and (c)). 
 This collective mode resonance appears in the $A_{1g}$ symmetry since the Raman response is screened by long range Coulomb interaction in this symmetry. 
 From Ref.\ \cite{Montiel15a}.}
\end{figure}

Emergent symmetries also have signatures in terms of collective modes.
In a recent work we argued that the $A_{1g}$-mode observed in Raman
scattering very close in energy to the neutron mode is such a signature
of the SC-CDW $SU(2)$ symmetry \cite{Montiel15a}, see Fig.\ \ref{fig:raman}.
The collective mode used in this work was associated with the $\eta$-
operator of Eqn.(\ref{eq:1}) with axial wave vector, thus associated
to the triplet representation Eqn.(\ref{eq:5-1-1}). The presence
of the two orders in conjunctions was needed in order for the Raman
scattering vertex not to vanish. The model could account for the absence
of observation of this order in the $B_{1g}$ and $B_{2g}$channels.

Inelastic neutron scattering has reported since the very early days
the presence of a collective mode in the underdoped regime, centered
around the AFM wave vector $Q=(\pi,\pi)$, and at a finite energy
around $E=41$meV for the compound YBCO \cite{Rossat,Hinkov04}. Many
theories, based on an RPA treatment of a magnetic spinon mode below
the SC gap have been produced in order to explain this very characteristic
feature of the cuprates \cite{Demler95,Demler04,Tchernyshyov01}.
The $SO(5)$ theory was originally devoted to the study of this mode
\cite{Demler95}. The RPA theories, reproduce successfully the position
of incommensurate signal around $Q=(\pi,\pi)$, having the typical
``hour-glass'' shape in the energy-momentum space. The present theories
have difficulties to account for the fact that this signal remains
inside the PG phase, changing form from the ``hour-glass'' to a
``Y'' shape, namely acquiring some extra low energy spectral weight
at $Q=(\pi,\pi)$. The proposed RPE state is an excitonic state with
excitations around a bunch of $2\mathbf{k}_{F}$- wave vectors in
the anti-nodal region. Thus it behaves a little bit as a ``charge
superconductor'', that in the simplest models, will gap out the electronic
degrees of freedom precisely as a superconductor would do. We believe
the RPE state can also account for the extra spectral weight at $Q=(\pi,\pi)$,
which will be presented in a future work \cite{Montiel16}.

\subsubsection{ARPES}

We turn now to angle-resolved photoemission spectroscopy (ARPES),
which have been very influential in our understanding of the PG phase
of these materials, especially with the seminal observation of Fermi
arcs in this phase . A recent ARPES experiment on Bi-2201 has been
very important in our understanding of the formation of the PG \cite{He11,Shen:2005ir}.
Dispersion cuts close to the Zone Edge show that the PG opens at a
typical momentum larger than the momentum relating the two Fermi points
$2\mathbf{k}_{F}$. Moreover when the dispersion cuts get closer
to the nodes, the PG closes from below rather than from above. It
has been argued that this set of peculiar features can only be explained
by a PDW state (a finite momentum SC state), since only the particle-hole
reversal specific to the pairing state can account for the closing
of the gap from below \cite{Lee14,Wang15a}. 
\begin{figure}[h]
\includegraphics[width=80mm]{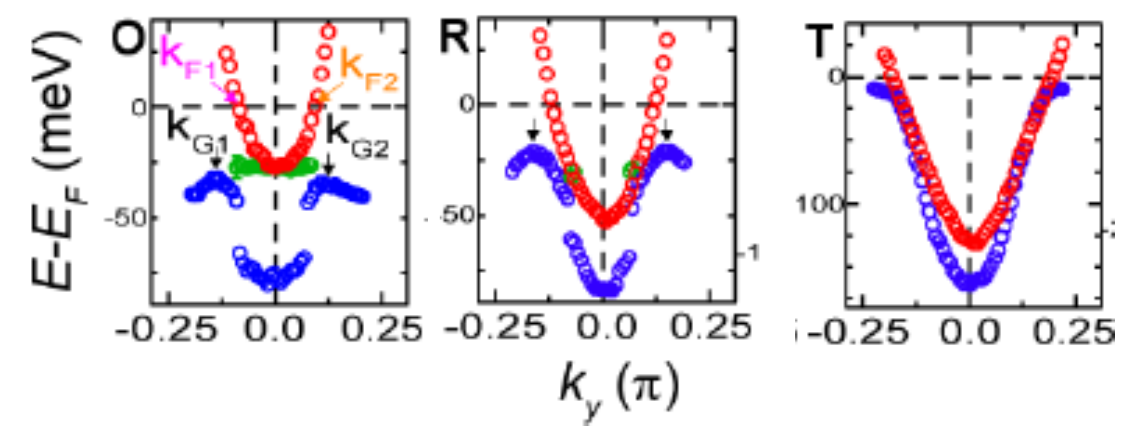} \vspace{-0.5ex}
 \caption{\label{fig:gapdisp} (Color online)
Electronic dispersion measured by ARPES in the superconducting (blue and green) and normal state (red).
Cuts taken at constant $k_x$: close to the zone edge (O)($k_x=\pi$), 
close to the nodal region (T, $k_x=0.6 \pi$) and in an intermediate case (R, $k_x=0.8\pi$).
The closing of the gap can be observed from the antinodal to the nodal zone.  
From Ref.\ \cite{He11}.}
\end{figure}
We argue that the RPE
state provides another explanation for this fascinating set of data.
Besides the multiplicity of the wave vectors, the key ingredient is
the non locality of the excitons. In particular, in the reciprocal
space, they form within a finite window in the anti-nodal region,
which can account for the natural closing in energy of the gap, both
from above and below (ARPES does not see the positive energies), so
that we have only to account for the negative part of the spectrum
\cite{Montiel16}.

\subsubsection{Loop currents}

The observation of a $\mathbf{Q}=0$ signal in neutron scattering
at a temperature line following $T^{*}$ \cite{Fauque06} is one of
the mysteries of the PG phase, which has been interpreted in terms of the formation
of intra-unit-cell $\Theta_{II}$-loop currents \cite{Gull:2013hh}. Although it
is commonly understood that a $\mathbf{Q}=0$ phase transition does
not open a gap in the electronic spectrum, and thus the 
$\Theta_{II}$-loop-current phase alone cannot be responsible for the origin of the
PG, any proposal for the PG phase has to account for the signal observed
in the neutron scattering experiment. 
\begin{figure}[h]
\includegraphics[width=50mm]{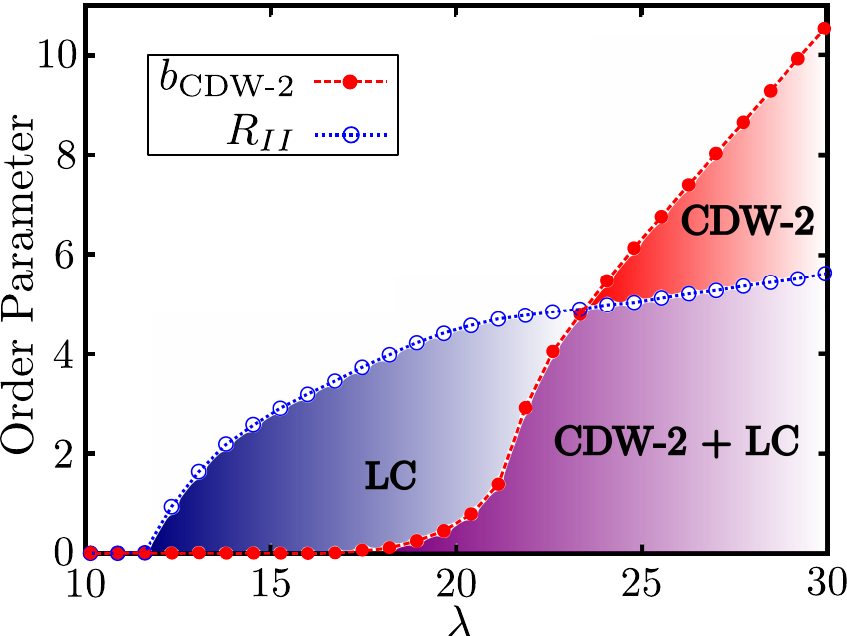} \vspace{-0.5ex}
 \caption{\label{fig:2dcdw} (Color online) Coexistence between $\Theta_{II}$-loop current order
 parameter and bidirectional $d$-wave CDW as a function of the interaction strength $\lambda$.
 From Ref.\ \cite{Carvalho15b}.}
\end{figure}
We have produced two studies
within the EHS model regarding the possibility of coexistence of charge
orders and loop currents \cite{Carvalho15,Carvalho15b}. In Ref. \cite{Carvalho15}, we have shown, within a saddle-point approximation, that
the $\Theta_{II}$-loop-current order cannot coexist with a d-wave CDW with 
diagonal wave vectors. As a result, we have offered this scenario as the possible reason, which explains why a d-wave charge order was never observed along the diagonal direction in the cuprates. 
In a subsequent work \cite{Carvalho15b}, we have demonstrated that a similar behavior is displayed by the d-wave CDW along axial directions described by uni-directional wave vectors
(i.e. of the stripe-type), since the $\Theta_{II}$-loop-current order is also detrimental to the latter order. By contrast, we have shown that bi-directional (i.e. checkerboard)  d-wave CDW and PDW
along axial momenta, which are in turn related by the emergent $SU(2)$ pseudospin symmetry pointed out previously, are compatible with the $\Theta_{II}$-loop-current order, since all these orders can coexist with one another in the phase diagram (see, e.g., Fig.\ \ref{fig:2dcdw}). These theoretical predictions agree, most spectacularly, with recent STM results {\cite{Hamidian16}}
and also with x-ray experiments \cite{Blanco-Canosa14}.

\subsubsection{Pump probe experiment}

A recent pump probe experiment also gives some evidence of the presence
of strong SC fluctuations at an intermediate energy scale \cite{Kaiser14}
in underdoped cuprates. In the first series of pump probe experiments
\cite{Averitt01,Pashkin10}, the cuprate was excited up to $1.5$
eV and relaxation at the pico-second scale - observed in the optical
THz regime, destroyed the Cooper pairs and showed two typical energy
scales, one related to the PG regime and one associated with the formation
of the coherence SC phase. Those two scales are typically the ones
observed, for example, in the $dI/dV$ response of STM microscope.
\begin{figure}[h]
\includegraphics[width=50mm]{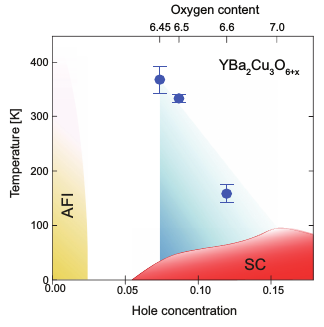} \vspace{-0.5ex}
 \caption{\label{fig:neqgap} (Color online) Schematic phase diagram for YBCO
proposed in Ref.\ \cite{Kaiser14}. Under out-of-equilibrium conditions
realized by optical pump-probes, a high mobility phase in the blue
shaded area can be realized that extends much above the critical temperature
of equilibrium SC. From Ref.\ \cite{Kaiser14}.}
\end{figure}
But in a recent experiment, the excitation was much weaker, in the
mid Infra red regime \cite{Kaiser14} which enabled to scan the properties
of the PG phase without destroying the Cooper pairs. What was found
resembles to a pico-second photograph of SC fluctuations with the
superfluid density $\rho_{s}\sim\omega\Delta\sigma_{2}$ shooting
up in the PG phase, up to temperatures of 300 K , see Fig.\ \ref{fig:neqgap}.
This pico-second ``photograph'' of the superfluid density was shown
to follow the PG temperature as a function of doping.

\subsubsection{General phase diagram}

A general look at the phase diagram of the cuprates singles out many
enduring mysteries, and one of the most enduring one is the linear-in-$T$ resistivity around optimal doping. This phase was identified
in a seminal work as a Marginal Fermi Liquid (MFL) \cite{Varma89,Varma97},
and it is still a challenge for theories to account for this phenomenon.
Recent in-depth experiments show a more complex behavior of the resistivity
with temperature, with a part linear in $T$ and a residual part going
like $T^{2}$ when the strange
metal regime is approached from the right hand side of the phase diagram 
\cite{Hussey:2008tw,Hussey:2011kp}. Two schools of ideas have been
advanced to explain this very unusual phenomenon. In the first school
of ideas, it is believed that the proximity to the Mott transition
creates a very strongly correlated electronic medium where the electron
mean free path is so weak that we are above the Ioffe-Regel limit
for the MFL regime \cite{Georges96}. The second school of ideas
advances that the resistivity slope as a function of temperature is
very steep, so that the second MFL regime extends far beyond the Ioffe-Regel limit
at low enough temperatures. Within this second viewpoint, the challenge
is to suggest a QCP beyond the dome which could produce a very resistive
scattering behavior for the conduction electrons. It is precisely
what the quantum critical version of the RPE state does. 
\begin{figure}[tbc]
\includegraphics[width=50mm]{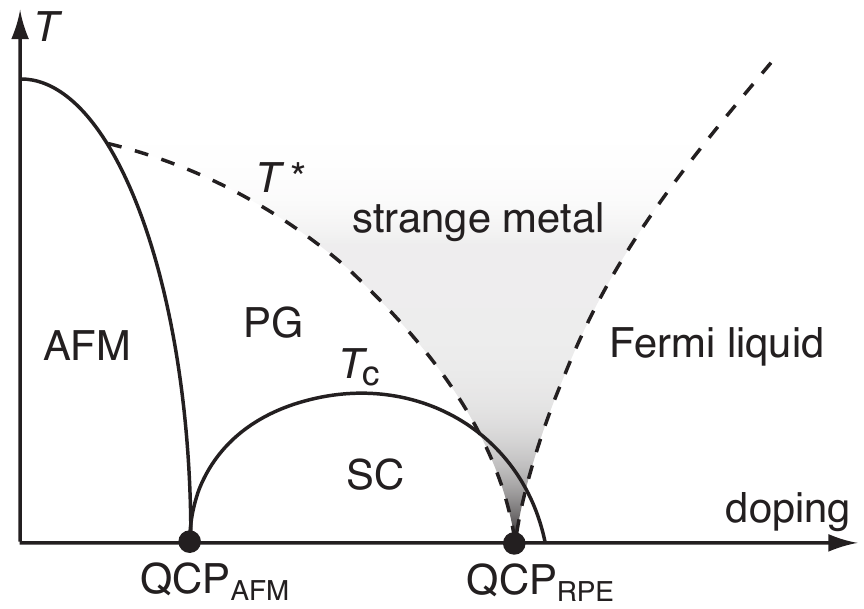} \vspace{-0.5ex}
 \caption{\label{fig:strangemet} (Color online) Schematic phase diagram of
cuprate superconductors with the RPE state, as proposed in \cite{Kloss15a}.
The quasi one-dimensional scattering in the vicinity of QCP$_{RPE}$
produces the resistivity and the electronic self-energy anomaly observed
in the strange metal phase. From Ref.\ \cite{Kloss15a}.}
\end{figure}
Electronic
scattering through quantum critical excitonic modes shows a quasi-one
dimensional behavior, each electron scattering preferentially through
its most favorable $2\mathbf{k}_{F}$ wave vector \cite{Kloss15a},
and produces a resistivity that behaves as $\rho\sim T/\log T$ within a Boltzmann semiclassical calculation and the electronic
self-energy that reads $\Sigma\left(i\epsilon_{n}\right)\sim i\epsilon_{n}/\log\left|\epsilon_{n}\right|$
in the ``Strange Metal'' (SM) regime (see Fig. \ref{fig:strangemet}). On
the same line of thought, maybe one of the most difficult feature of
the PG to account for in any theory is that it is insensitive to a large
amount of $Zn$-doping or irradiation by electrons \cite{Alloul09,RullierAlbenque:2000fl,RullierAlbenque:2011ji},
which locate on $Cu$ sites and produce strong disorder which exclude
the doped $Cu$-site in the unitary limit \cite{Pepin98}. The $T^{*}$-line
is not affected and also the slope of the resistivity in the SM regime
does not change \cite{RullierAlbenque:2000fl}. It is difficult for
any state of matter to have the sufficient robustness to show no sensitivity
to such a strong perturbation. One way the RPE state could resist
is through the non-locality of the excitons (i.e. the particles-hole pair), which
can typically being created at site $\mathbf{r}$ and removed at site
$\mathbf{r'}$ with the typical correlation $\left\langle c_{\mathbf{r}}^{\dagger}c_{\mathbf{r'}}\right\rangle $
\cite{Montiel16}.

\section{Conclusion}
\label{conclusion}

To conclude, within the two large views of the cuprates where, on
one hand, the physics of the PG is solely determined by strong correlations
and the proximity to the Mott transition, and the other view where
the qualitative features of the physics of the PG are governed by
some hidden symmetry, the present work makes a clear discrimination
in favor of the latter. It is claimed here that the physics of
the PG and the SM phase are controlled by an emergent $SU(2)$ symmetry.
Many properties of the underdoped cuprates can be captured within
the pseudo-spin theory, the non-linear $\sigma$-model associated
to this symmetry and the pseudo spin-flop physics between the SC and
charge channel. We also claim that $SU(2)$ superconducting fluctuations
proliferate at intermediate energy scales in the physics of these compounds, and 
are are the key ingredient to understand the PG phase. At lower energy,
they lead to the formation of the RPE state, which we believe has
a lot of promising features to be the solution for the PG. At even
lower temperatures, the $U(1)$ phase fluctuations enter the game and
produce coherent axial CDW mediated by a Leggett-mode. The $SU(2)$
symmetry is present in the background of the whole underdoped region,
and it is expected that pseudo-spin partners of the various orders (such as the PDW partner of the CDW order) have recently been observed experimentally \cite{Hamidian16}. 
Lastly, what is the influence of the Mott transition on the
phase diagram of the cuprates ? We believe it will qualitatively affect
the physics up to roughly 6\% of doping. Below 6\% of doping, techniques
adequate to describe the very strongly coupled regime will capture
the physics \cite{LeTacon06,Gull:2013hh,Sordi2012} . Beyond 6\% doping,
the physics is qualitatively protected by the $SU(2)$ symmetry. A
very revealing experiment is the variation of the nodal velocity $v_{\Delta}$
as a function of doping extracted from ARPES data (Ref. \cite{Vishik12}). A tri-sected dome is observed
with three distinct regimes, 1) below 6\% doping, 2) between 6\%
and 19\% of doping and 3) above 19 \% of doping. Within the $SU(2)$
theory, as with all theories controlled by an emergent symmetry,
the critical value of 6\% of doping is precisely the point where the
Mott physics becomes dominant. Within the strongly correlated viewpoint,
the typical doping of 6\% is difficult to interpret.

\begin{figure}[h]
\includegraphics[width=50mm]{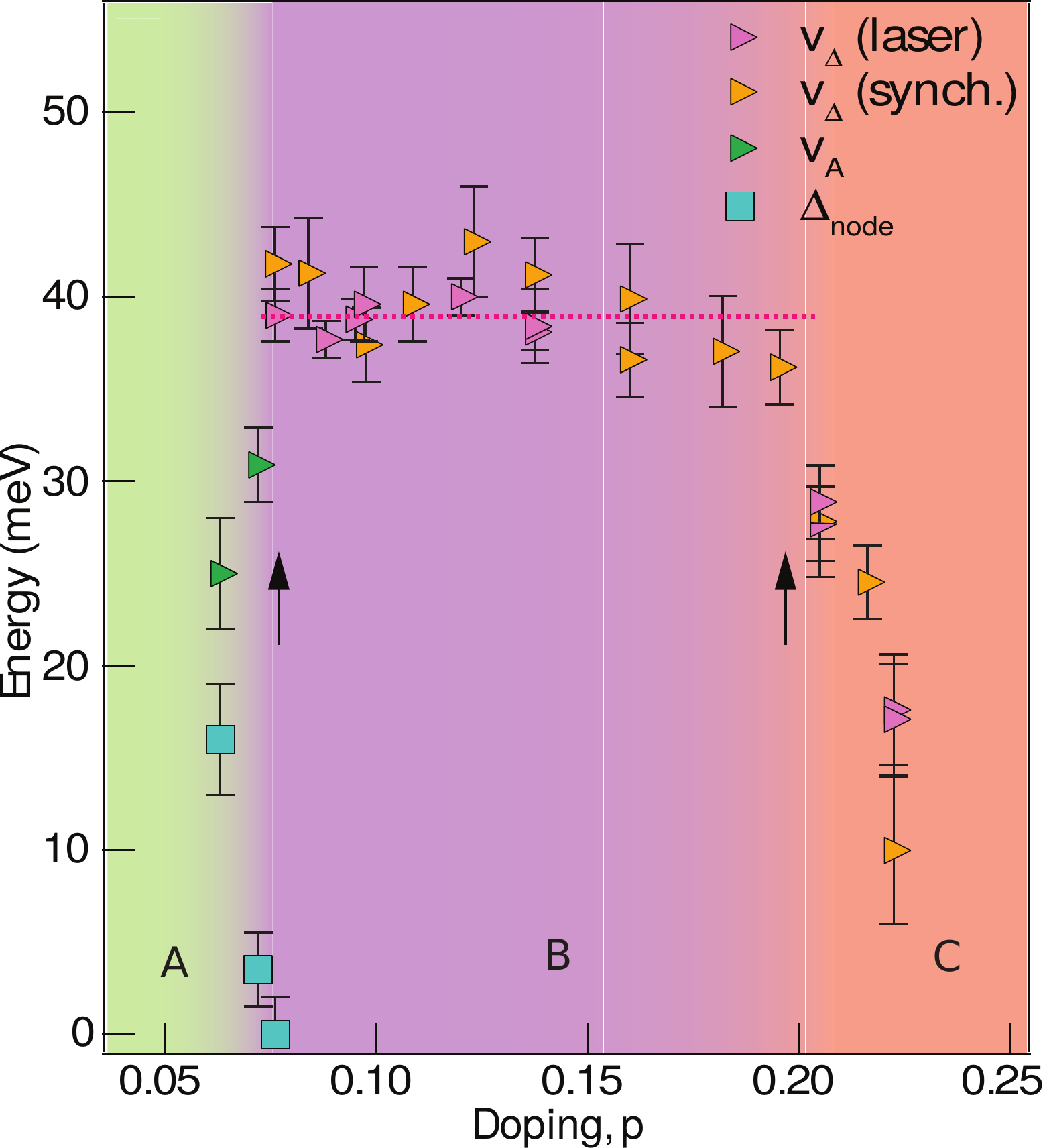} \vspace{-0.5ex}
 \caption{\label{fig:tstararpes} (Color online) 
ARPES experiments on BSCO performed in Ref.\ \cite{Vishik12}. 
The doping dependence of the gap velocity $v_{\Delta}$ reveals 
three distinct regime: two regions at low and high doping where $v_{\Delta}$ drops and a third regime in-between, where $v_{\Delta}$ reaches a plateau.
From Ref.\ \cite{Vishik12}.}
\end{figure}

\begin{acknowledgments}
We are grateful to H.\ Alloul, Y.\ Sidis, P.\ Bourges and A.\ Ferraz for stimulating and helpful discussions. 
This work was supported by LabEx PALM (ANR-10-LABX-0039- PALM),  ANR project UNESCOS ANR-14-CE05-0007, as well as the grant Ph743-12 of the COFECUB. CP and TK also acknowledge hospitality at the IIP (Natal, Brazil) where parts of this work were done.
\end{acknowledgments}

\bibliographystyle{phaip}
\bibliography{Cuprates}

\end{document}